\documentclass[11pt,3p,review,authoryear]{elsarticle}

\journal{International Journal of Forecasting}
\biboptions{longnamesfirst}

\usepackage[colorinlistoftodos]{todonotes}

\usepackage{graphicx}%, epstopdf}
\usepackage{float}
\usepackage{pgfplots}
\RequirePackage{amsmath,amssymb,amsthm,dsfont}
\usepackage{listings}
\usepackage{array}
\usepackage[caption=false,font=footnotesize]{subfig}
\usepackage{pdflscape}
\usepackage{comment}
\PassOptionsToPackage{hyphens}{url}\usepackage{hyperref}
\newcommand{\I}{\mbox{${\mathbf I}$}}

\newcommand{\bX}{\mbox{${\mathbf X}$}}

\newcommand{\cT}{\mathcal{T}}
\newcommand{\cS}{\mathcal{S}}

\begin{document}

\begin{frontmatter}

\title{Hierarchical transfer learning with applications for electricity load forecasting.}

\begin{abstract}
The recent abundance of data on electricity consumption at different scales opens new opportunities and highlights the need for new techniques to leverage information present at finer scales in order to improve forecasts at wider scales. In this work, we take advantage of the similarity between this hierarchical prediction problem and transfer learning where source data are observed at a low aggregation level and target data at a global level. We develop two methods for hierarchical transfer learning, based respectively on the stacking of generalized additive models and random forests (GAM-RF). We also propose and compare adaptations of online aggregation of experts in a hierarchical context  using quantile GAM-RF as experts. We apply these methods to two problems of electricity load forecasting at the national scale, using smart meter data in the first case, and regional data in the second case. For these two usecases, we compare the performances of our methods to that of benchmark algorithms, and we investigate their behavior using variable importance analysis. Our results demonstrate the interest of both methods, which lead to a significant improvement of the predictions.
\end{abstract}


\end{frontmatter}

%%%%%%%%%%%%%%%%%%%%%%%%%%%%%%%%%%%%%%%%%%%%%%%%
%%%%%%%%%%%%%%%%% INTRODUCTION %%%%%%%%%%%%%%%%% 
%%%%%%%%%%%%%%%%%%%%%%%%%%%%%%%%%%%%%%%%%%%%%%%%
\section{Introduction}

The recent abundance of electricity consumption data at low aggregation level, due in part to the development of smart meters, opens up many prospects for electricity consumption forecasting (e.g. see \cite{Wang2019}). However, with these new perspectives come new challenges, among which is the question of how to include these data obtained at a finer scale (corresponding to a household or to a smaller geographical area), which can be used to create forecasts at a fine scale, into a prediction at a wider scale (at the national scale for example). In this work, we present two methods for leveraging our ability to predict a variable of interest at a finer scale, with the goal of exploiting these predictions to improve prediction at a larger scale. This problem, which consists in taking advantage of the similarities (e.g. similar dependency to explanatory variables, common drift in distributions) existing between forecasting problems at different scales, can be naturally formulated in the framework of transfer learning. 

Transfer learning methods aim at transferring the knowledge acquired from solving given problems (referred to as \textit{source} problems $\cS$) to address an other problem of interest (referred to as the \textit{target} problem $\cT$). In a supervised predictive machine learning setting, the objective is to predict a variable of interest $Y^{\cT}$ using covariates $\bX^{\cT}$. To do so, the learner relies on a set of observations $(\bX_i^{\cT}, Y_i^{\cT})_{i \in \{1, \dots, n^{\cT}\}}$, drawn from a joint distribution $\mathbb{P}^{\cT}$. Popular methods rely on minimizing the empirical risk corresponding to a given loss function over a set of possible learners (e.g. tree based, neural nets, generalized additive models). When the learners and the loss are chosen appropriately, the estimated model will have good forecasting accuracy on a new dataset as long as the size of the training set is sufficiently large, and the marginal and joint distributions remain unchanged in the test set. In a wide range of real-world applications, these conditions are not satisfied. Classical examples of such situations include tasks requiring massive training sets, such as computer vision or natural language processing. When dealing with temporal data, one may be confronted with changes in the distribution leading to large prediction errors over the prediction period. If we consider that the historical data correspond to a first problem, and the prediction period to a different but related problem, we are faced with the following challenge: we must exploit the similarity between the two tasks, in order to take advantage of the abundance of historical data, while ensuring adaptability to the new task.

Transfer learning aims to tackle such a problem. It has attracted increasing attention in machine learning and has been used in many applications (see  \cite{olivas2009handbook}). In many practical situations, a relatively small quantity of data from the target distribution $\mathbb{P}^{\cT}$ is available. In some cases, one also has access to a larger dataset, with a different distribution denoted $\mathbb{P}^{\cS}$, which can be used to solve a task related to the target problem.  A key assumption is that $\mathbb{P}^{\cT}$ and $\mathbb{P}^{\cS}$ are related in a way that can be leveraged by the transfer learning method. These distributions may be defined on the same domain (in this case, the transfer learning problem is said to be homogeneous), or on different domains, in which case more complex transformations need to be developed (the transfer learning problem is then said to be heterogeneous). In this paper, we will focus on heterogeneous transfer problems, with differences between source and target in terms of features spaces, feature marginal distribution, and joint distribution.  

Surprisingly, although transfer learning is very popular in computer vision and text mining (see \cite{pan2010survey} and \cite{zhuang2020comprehensive}  for a survey), very few developments can be found in the time series forecasting community. In \citep{laptev2018applied} the authors fine-tune a pre-trained neural network using a large data set of individual electricity loads as source and some independent individual data as target. In \citep{capezza2021additive} additive stacking (an interpretable aggregation of experts)  is proposed combining models at the individual level and global one for probabilistic forecasting of individual demand. In \citep{obst2021adaptive, obst2021transfer} the authors propose a fine-tuning approach as well  as online updates to transfer information from Italian Data to French Data in order  to  improve electricity load forecasts during the COVID lockdown. Here the transfer is both in time (from past data (source) to future data (target)) and space (from one country to another).

Hybridizing statistical models with modern machine learning tools recently proved to be an efficient strategy to forecast time series (see \cite{SMYL202075} top rank at the M4 competition). \cite{anderer2022hierarchical} propose to transfer time-series features for bottom-up forecasting in the context of M5 competition with intermittent time series at the low aggregation level. N-BEATS, a deep learning forecasting approach is proposed at the global level and boosting trees with LightGBM at the bottom level. In the present paper, we develop new transfer learning methods for hierarchical prediction, that leverage data available at a fine scale to improve prediction at a wider scale. The first approach, presented in more detail in Section \ref{subsec:stackedGAMRF}, is based on the design of new features learned from the source data. These features are then used as input in a Random Forest (RF) stacked with a Generalized Additive Model (GAM). Stacking an ensemble of forecasting models for time series forecasting has already been done successfully (see e.g. \cite{khairalla2018short, moon2020combination} for load forecasting, \cite{dong2021wind} for wind power forecasting, \cite{xenochristou2020ensemble} for water demand forecasting and \cite{zhai2018development} for pollution forecasting). Previous work usually combined the forecasts in a meta-learner whereas we propose to combine features from GAM with original covariates in the RF. This is new and allows to detect interactions not modeled in GAM or appearing online (typically interaction with time in the context of drift). Our empirical results suggest that the feature design method presented above, combining stacked GAM and RF, allows to improve prediction at an wider scale by leveraging knowledge acquired from data at a finer scale. Unfortunately, this approach relies on knowledge acquired on a training set, which may not be relevant if a change in distribution occurs during the test period. For this reason, these methods are not adaptative to a brutal change in distribution both at the fine and the wider scale, and to leverage the relationship between both. To ensure adaptivity in our model, we propose a second transfer learning approach based on online aggregation of experts. In the hierarchical context, \cite{goehry2019aggregation} and \cite{bregere2021online} show that aggregating experts designed on different nodes of a hierarchical partition of the data (statistical clustering based on temporal or exogenous information, spatial partition) improves forecasting performances compared with classical bottom-up approaches. Our online aggregation strategies leverage similarities between shifts in distribution at the local scale and at the global scale, in order to adapt more quickly. We also present a new way of designing relevant experts in this context.

\subsection{Contributions and outline of the paper}
In this paper, we develop two methods for leveraging information available at a fine scale to improve prediction at a wider scale, based respectively on \textit{feature design} combined using \textit{stacked generalized additive models and random forests}, and on \textit{online aggregation of experts}. These methods are presented in Section \ref{sec methodology}, and illustrated on two real-world problems. In Section \ref{sec:smartmeter}, we apply the first method to the problem of electricity load forecasting at the national level, relying on smart meter data. In Section \ref{sec:Covid}, we combine these methods to obtain adaptative methods for forecasting electricity consumption at the national level during the Covid-19 pandemic period, using data available at the regional level. We demonstrate the interest of our proposed approach in both cases. Our results indicate that both the stacking of GAM and RF and the use of features designed on data at finer scale lead to improvements in the forecasts at wider scale. Moreover, the use of multi-scale information transfer through aggregation of experts also increases the quality of wide-scale forecasts. Strikingly, our results indicate that in the two usecases, the presented methods can improve wide-scale predictions by using fine-scale predictions, even when no hierarchical constraints are implemented.

For reproducibility of the results, please find the code and data in the supplementary material: \url{https://drive.google.com/file/d/1hdCEHKpVXt6zoSi7n7xEA0oUKW-_uEeD/view?usp=sharing}.

%\url{https://github.com/SolenneGaucher/Hierarchical-Transfer-Learning-Covid19.git}.

\section{Concepts and algorithms}

In this section, we present the different statistic tools composing our transfer learning approach: generalized additive models, (quantile) random forests, and online aggregation of experts.

\subsection{Generalized additive models}

Generalized additive models \citep{wood2006generalized} are a simple class of models that model a response as a sum of smooth non-parametric functions of the covariates. Partially linear additive models \citep{amato2017estimation}, which are a special case of generalized additive nonparametric models (GAM), retain the parsimony and interpretability of linear models and the flexibility of nonparametric additive regression, by allowing a linear component for some predictors which are presumed to have a strictly linear effect, and an additive structure for other predictors. This choice of both linear and non-parametric components allows to reduce the degrees of freedom and to mitigate the problem known as ``curse of dimensionality''.

Given  observations  $\{(Y_t, \mathbf{X}_t^{(1)}, \mathbf{X}_t^{(2)})\}_{t=1}^n$, where $Y_t$ is the response at time $t$, $\mathbf{X}^{(1)}_t=(X_{t,1}^{(1)}, \dots, X_{t,d_1}^{(1)})^T$ and $\mathbf{X}^{(2)}_t=(X_{t,1}^{(2)}, \dots, X_{t,d_2}^{(2)})^T$ are vectors of covariates, the partially linear generalized additive model assumes that
\begin{equation} \label{plmm}
Y_t = b+\left(\mathbf{X}_{t}^{(1)}\right)^T\boldsymbol{\beta} + \sum_{j=1}^{d_2} f_j (X_{t,j}^{(2)}) + \epsilon_t, \quad t=1,\dots, n,
\end{equation}

where $b$ is the intercept, $\boldsymbol{\beta}$ is the $d_1 \times 1$ vector of unknown coefficients for linear terms,  $f_j$  are unknown nonlinear real valued components and the $ \epsilon_i$'s  are i.i.d random variables with mean 0 and variance $\sigma^2$ independent of the covariates. In order to ensure that the model is identifiable, one requires that the linear covariates are centered and that identifiability conditions $\int f_j(t) dt=0$, $j=1,\dots, d_2$ hold. For the sake of simplification, and with some abuse of definition, such PLAM models will be referred hereafter as GAMs, and we will denote by $f_k(X_k)$ the effect of variable $X_k$, be it linear or non-parametric.

Such models, together with procedures that achieve estimation and simultaneous consistent variable selection, have proven their ability to cope with high-level aggregate electricity data in previous work: \cite{Goude2013} applied it to french substations consumption and \cite{FanHyndman2012} show their interest for regional Australia's load forecasting. Moreover they can be applied efficiently to forecast electricity data at different levels of aggregation \citep{amato2021forecasting}. In the following, GAMs are trained in R using the library mgcv \citep{GAMbook}.

\subsection{Random forest and quantile regression forest}

Random forests (RF) are a powerful black box approach for modeling complex regression relationships (see \cite{breiman2001random}). The very general underlying  model behind a random forest regression assumes that $y_t = h(X_t)  + \varepsilon_t$, where $g$ is a generic, non-parametric function, and $\varepsilon_t$ is an independent gaussian noise. Because of the generality of the model, RF necessitates very little prior knowledge of the problem. RF are obtained by aggregating an ensemble of base learners generated by applying classification and regression trees (CART, see \cite{breiman1984cart}) on different subsets of the data obtained with bagging and random sampling of covariates. One nice feature of random forests is  that they can be easily used for quantile regression as presented in \cite{meinshausen2006quantile}.

In our applications, we use the procedure {\tt ranger()} from the R toolbox {\tt ranger} for the random forest fits. The default parameters are used ($500$ trees, $mtry=\sqrt{p}$, unlimited tree depth). In future work, these values could be optimized  in a more refined way by combining {\tt ranger} with procedures from the R library {\tt caret}, at the cost of increasing CPU time.

\subsection{Online aggregation of experts}
Online robust aggregation of experts \citep{Cesa-Bianchi:2006} is a powerful model agnostic approach for time series forecasting. It consists in combining in a streaming fashion different forecasts (called experts) according to their past performances. When experts forecasting a variable of interest at a finer scale are aggregated so as to forecast this variable at the wider scale, this allows to transfer knowledge between these different scales. Aggregation of experts was recently applied in a forecasting competition (see \cite{Farrokhabadi2021}), where 2 of the 3 first teams (see \cite{de2021state}, \cite{ziel2021smoothed}) applied these approaches to forecast electricity load consumption during the COVID lockdown in a big city (unknown localization). In this changing context, online aggregation allows to adapt to changes in distributions and to track the performance of the best expert.

We propose here a short description of sequential expert aggregation for forecasting. A complete presentation of these methods can  be found in \cite{Cesa-Bianchi:2006}. Sequential expert aggregation assumes that data are observed sequentially: the target variable (here the  electricity consumption) is supposed to be a bounded sequence $Y_1,\dots,Y_T \in [0,B], B>0$, which we want to forecast step by step for every time $t$. At each time $t$, $N$ experts provide forecasts of $Y_t$, denoted $\left(\hat{Y}_{t}^{1},\dots,\hat{Y}_{t}^N\right) \in [0,B]^N$. These experts can come from a statistical model, a physical model, or expert advice projection. The aggregation algorithm chooses weights $\hat{p}_{j, t} \in \mathbb{R}^N$, and returns as forecast for $Y_t$ a weighted average $\hat Y_t =  \sum_{j=1}^N \hat{p}_{j, t} \hat{Y}_{t}^j$ of the $N$ forecasts. Then, $Y_t$ is observed and instance $t+1$ begins. In the following, we will consider only convex aggregation (with weights $\hat{p}_{j, t}$ summing to one and in $[0,1]$).

The performances of experts and aggregation forecasts are evaluated according to a convex loss function. We will consider here the square loss $\ell_t(x) = (Y_t-x)^2$. At time $t$, expert $k$ suffers loss $\ell_t(\hat{Y}_{t}^k) = (Y_t - \hat{Y}_t^k)^2$ and the aggregation $\ell_t(\hat Y_t) = (Y_t - \hat Y_t)^2$. The purpose of expert aggregation is to minimize the total loss $\sum_{t=1}^T (Y_t - \hat Y_t)^2$ that can be expressed:
\[
%	\label{eq:error}
	\frac{1}{T} \sum_{t=1}^T (Y_t - \hat Y_t)^2 \quad \triangleq \quad \frac{1}{T} \sum_{t=1}^T (Y_t - \hat Y_t^{\ast})^2 + R_T \,,
\]
$\hat Y_t^{\ast}$ is called an oracle and can be viewed  as an optimal (unknown before the forecasting run) forecast. $R_T$ is the regret term corresponding to the error suffered by our algorithm relatively to the error of the oracle. Some algorithms are proposed to achieve low regrets. In our study, we use the ML-Poly algorithm proposed in \cite{gaillard2014second} and implemented in the \texttt{R} package \texttt{opera} \citep{gaillard2016opera}. This algorithm tracks the best expert or the best convex aggregation of experts by giving more weight to an expert that will generate a low regret. This makes this algorithm particularly interesting as no parameter tuning is needed.

\section{Hierarchical stacking} \label{sec methodology}

In this section, we present our methodological contributions.
The first one relies on learning new features using data from the source distribution. These features are then used as input in a stacked GAM and RF model. The second one is to design new aggregation strategies to adaptively forecast variables on a bi-level hierarchy.

subsection{Feature design for stacked GAM and RF}\label{subsec:stackedGAMRF}

\paragraph{Features design using the source data} In the following, we suppose to have access to two data sets 
$$\mathcal{D}_{\cT}=\left(\bX_{t}^{\cT}, Y_{t}^{\cT}\right)_{t=1, \ldots, n_{\cT}}\quad \hbox{ and } \quad \mathcal{D}_{\cS}=\left(\bX_{t}^{\cS}, Y_{t}^{\cS}\right)_{t=1, \ldots, n_{\cS}},$$
where $\mathcal{D}_{\cT}$ is the target data set in the sense that the final objective is to forecast $Y_t^{\cT}$, and has underlying distribution $\mathbb{P}^{\cT}$ . $\mathcal{D}_{\cS}$, with underlying distribution $\mathbb{P}^{\cS}$, is an auxiliary source data set sharing some common properties with $\mathcal{D}_{\cT}$. We then want to exploit $\mathcal{D}_{\cS}$ in order to improve the forecast of $Y_t^{\cT}$.

While, in general, the covariates from the source and target datasets $\bX_{t}^{\cS}$ and $ \bX_{t}^{\cT}$ may belong to spaces of different dimensions, we may assume without loss of generality that there exists a subset $C$ of covariates that are common to both data sets. In the setting of electricity consumption forecasting, these common variables can be, e.g., calendar variables, or meteorological variables (at finer scale in $\mathcal{D}_{\cS}$, and at wider scale in $\mathcal{D}_{\cT}$). It is then natural to assume that these features will have similar effects on the variable of interest $Y_t$ in the target and source data set. To exploit this idea, we propose to learn the effect $f_k$ of a common feature $X_{t,k}$ such that $k \in C$, using the source dataset $\mathcal{D}_{\cS}$. More precisely, we fit a GAM on the dataset $\mathcal{D}_{\cS}$, and we extract the smooth function $f_k$ corresponding to the effect of covariate $X_{t,k}$. We then use $f_k$ to generate a new feature $f_k(X_{t,k}^{\cT})$, which we include in the target dataset $\mathcal{D}_{\cT}$. When the functions $f_k$ are learned from an auxiliary dataset $\mathcal{D}_{\cS}$ corresponding to observations at a finer scale, adding the corresponding features to the dataset $\mathcal{D}_{\cT}$ of observations at the wider scale allows to transfer knowledge in a hierarchical fashion.

Note that we can also use this technique to learn the effects of the covariates directly on the target dataset $\mathcal{D}_{\cT}$, and use them as new features in the regression. If we use one type of learner (e.g., GAM) to learn the feature, and we combine these features using a different type of learner (e.g., RF), we can hope to take advantage of both types of learners. This idea motivates the stacking of generalized additive models (GAM) and random forests (RF) presented below.

\paragraph{Stacked GAM and RF} Our models for the target problem are obtained by stacking GAMs and the correction provided by random forest regression trained on the target dataset.

GAM provide interpretable models and a natural way to incorporate expert knowledge into a statistical model. In addition, because of the smoothness assumptions imposed on GAM functionals, GAMs provide a good representation of the effects of important features and they can extrapolate out of training data. However, they only model the influence of pre-specified covariates or pairs of covariates, and can therefore fail to account for some interactions between inputs.

By contrast, Random forests (RF) can model complex regression relationships (see \cite{breiman2001random}): their black box design can capture well complex non-linear interactions. By definition, RF predictions are restricted to the convex hull of the outcomes $Y_t$ of the training data (all the possible mean of $y_i$). This behavior prevents them from producing aberrant predictions caused by extrapolation, even when trained on very small data sets, as can typically be the case in a transfer learning framework with a small target dataset and a high number of covariates \citep{balestriero2021learning}. To have the best of both worlds, we propose to stack these two approaches. Note that using other black box machine learning methods such as neural nets or boosting trees could be a potential improvement for future research.

The stacked GAM and RF algorithm (GAM-RF in  the following) consists in three steps:

\begin{enumerate}
  \item We first fit a GAM model as Equation \eqref{plmm} on the source data $\mathcal{D}_{\cS}$. We use the estimated GAM features $\left(f_k\right)_{k \in C}$ to create new features $\left(\left(f_k(\bX_{k,t}^{\cT})\right)_{k \in C}\right)_{t=1, \ldots, n_{\cT}}$ for the target dataset.
 \item We compute estimates of forecasting residuals on the target dataset $\mathcal{D}_{\cT}$ (either by cross-validation, block cross-validation or forecasting errors in an  online forecasting setting) denoted $\widehat{\varepsilon}_t$.
  \item We then fit a RF model on the augmented target dataset $\left(\widehat{\varepsilon}_t, \bX_{t}^{\cT}, \left(f_k(\bX_{k,t}^{\cT})\right)_{k \in C}\right)_{t=1, \ldots, n_{\cT}}$ to predict the GAM residuals $\widehat{\varepsilon}_t$. The final forecasts are obtained by summing GAM forecasts and the corrections provided by the RF.
\end{enumerate}

The method presented above allows transferring information through the new features $f_k$, used as input in the RF. In Section \ref{sec:smartmeter}, we illustrate this methodology by applying the stacked GAM-RF to predict electricity load at the national level for the United Kingdom, by leveraging data available at a finer scale collected by smart meters.

\subsection{Online aggregation of stacked experts for a bi-level hierarchy}
\label{ssec:online_hier}

Our experiments in Sections \ref{sec:Covid} and \ref{sec:smartmeter} show that GAM and RF stacking can exploit the data in the source dataset to improve prediction on a related target dataset. However, this method relies on the assumption that the source and target distributions are constant, and thus may not be robust if a change occurs in these distributions. In many hierarchical prediction situations, it is natural to assume that changes in the distribution at the least aggregated level (i.e., on the source data) and at the most aggregated level (i.e., on the target data) are related. One may then want to take advantage of the data available for the source problem to learn these changes more quickly and obtain more adaptive forecasts for the target problem. To achieve this goal, we propose to use online aggregation of quantile experts. We consider a hierarchical forecasting setting where the data $y_t$ are observed at a global level and in $K$ zones $y_{z,t}$ such that $y_t=\sum_{z=1}^K y_{z,t}$. We  denote $y_t^{norm}$ (resp. $y_{z,t}^{norm}$) the  normalized load at the global (resp. zonal) level. This normalization consists in dividing these time series by their empirical mean (computed on the source set). We propose different original ways to create \textit{experts} that will be aggregated online.

\subsubsection{Experts} \label{subsec:experts}

Online expert aggregation leverages the diversity of predictions made by different experts by combining their predictions. To obtain a diverse set of experts, we train our models to predict different quantiles of the target distribution. Designing experts for low and high quantiles present several advantages. On the one hand, these experts, when aggregated online to track the changes in the distribution of the load using convex aggregation, are particularly relevant since there is a high probability that the real consumption falls in the convex hull of the quantile experts. On the other hand, by doing so, we obtain experts with similar behavior across regions, that can share weights between the different regions and at the national level.  Indeed, it is reasonable to assume that when an expert receives a low weight in one region, it must receive a low weight in all regions. For example, in Section \ref{sec:Covid} we study the problem of electricity load forecasting in France at the national level using regional data. Measures taken in response to the Covid-19 epidemic in France resulted in a decrease in electricity load throughout the country: this change would correspond to low-quantile experts for the different regions receiving higher weights in the aggregation. Considering a vectorial aggregation model allows us to take advantage of the similar behavior of the quantile experts across regions.

In addition, we increase the diversity of the set of experts by considering different models to predict these quantiles, which we now describe. The stacked GAM and RF presented in Section \ref{subsec:stackedGAMRF} can be computed for each zone. We propose two ways of computing  stacking: one considering each zone individually (individual GAM-RF), and the other considering common models for zones and at the global level (common GAM-RF). We the following experts.

\begin{itemize}
    \item \textbf{GAM:} a GAM is fitted on each zone and at the global level on the normalized data resulting in $K+1$ scaled experts.
    
    \item \textbf{Individual GAM-RF:} using quantile  regression forests, we fit for each zone and at the global level 5 RF on the residuals of the GAM model. These RF predict quantiles at levels 0.05,  0.1, 0.5, 0.9, 0.95. By stacking the prediction of these RF and of the GAM, we obtain 5 scaled experts for each zone.
    
    \item \textbf{Common GAM-RF:} using quantile  regression forests, we fit 5 RF on the aggregated residuals for all zones and the global level. These RF predict quantiles at levels 0.05,  0.1, 0.5, 0.9, 0.95. By stacking the prediction of these RF and of the GAM, we obtain 5 scaled experts for each zone.
\end{itemize}

\begin{figure}[H]
  \begin{center}
    \includegraphics[width=\textwidth]{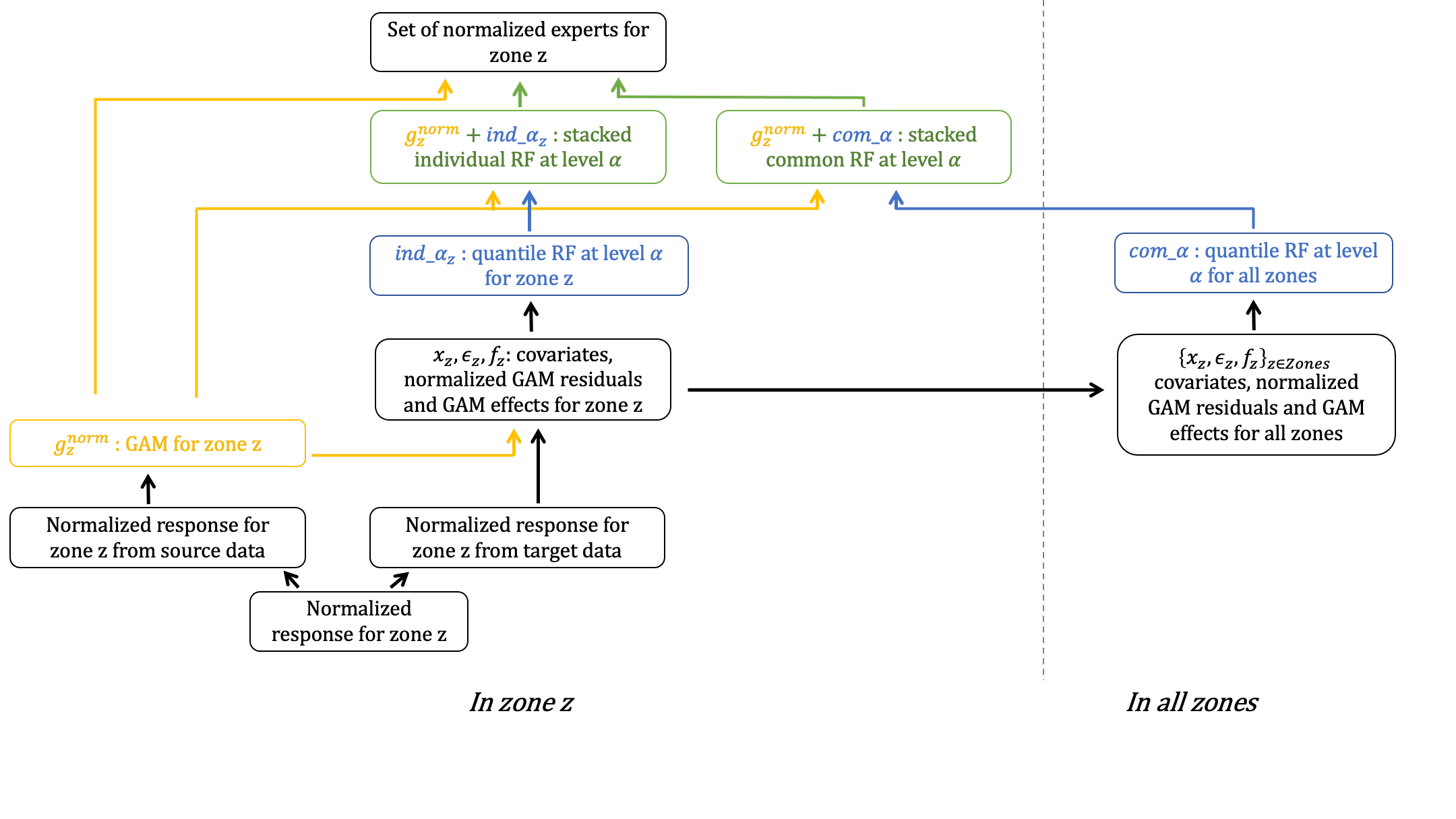}
    \end{center}
 \caption{\label{fig:stackedRF} Experts used for predicting the normalized responses for the different zones and at the global level.}
\end{figure}

These approaches are illustrated in Figure \ref{fig:stackedRF}. Quantile experts can be considered are possible scenarios of evolution of the data distribution that we will try to track online in the aggregation. Common RF are used to improve the transfer efficiency and capture common dynamics between  the zones.

\subsubsection{Aggregation strategies}\label{subsubsec:aggregation}

\begin{figure}[H]
  \begin{center}
    \subfloat[Fully disaggregated model.]{
    \includegraphics[width=\textwidth]{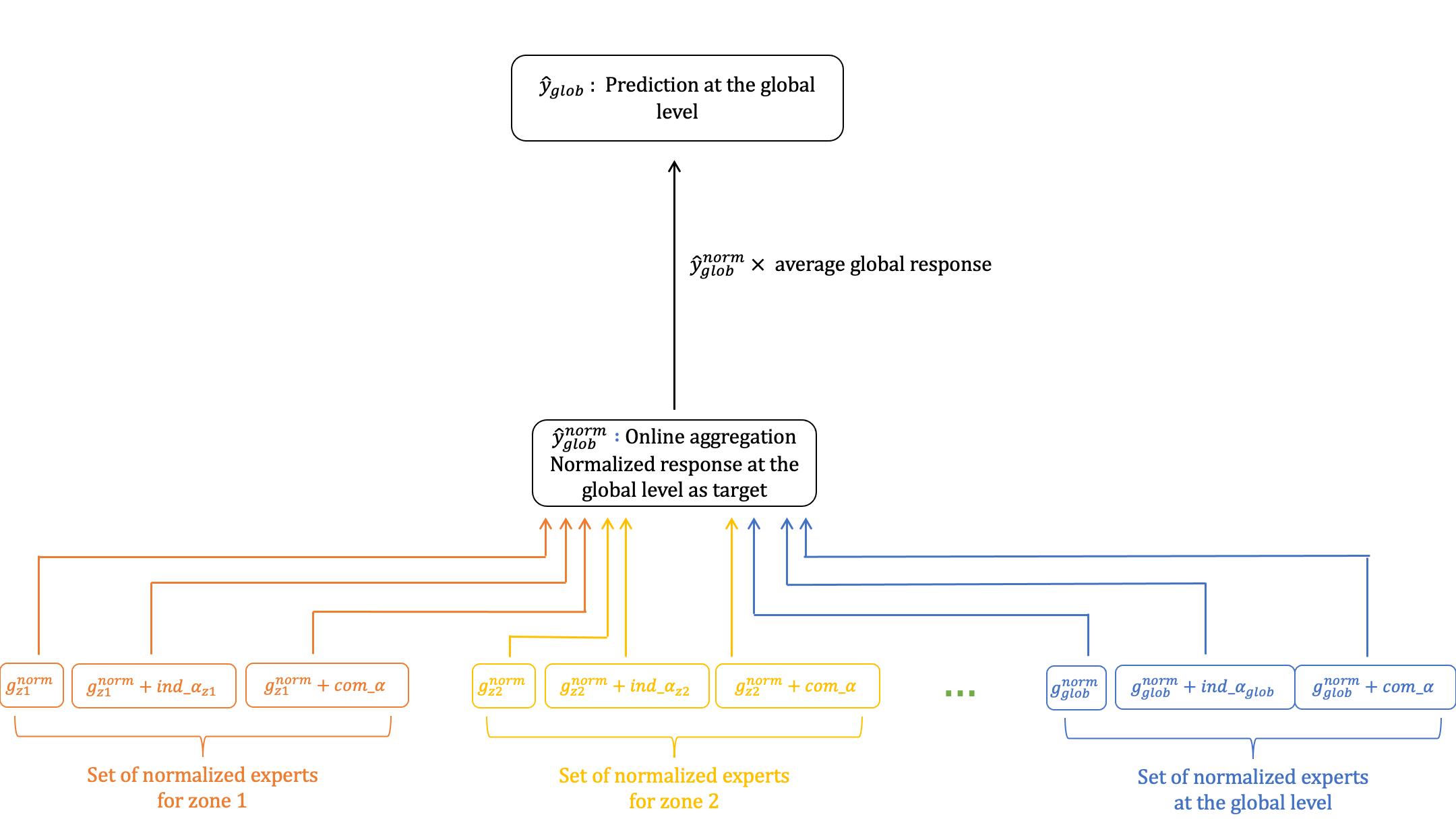}
                         }
                         
    \subfloat[Vectorial aggregation.]{
    \includegraphics[width=\textwidth]{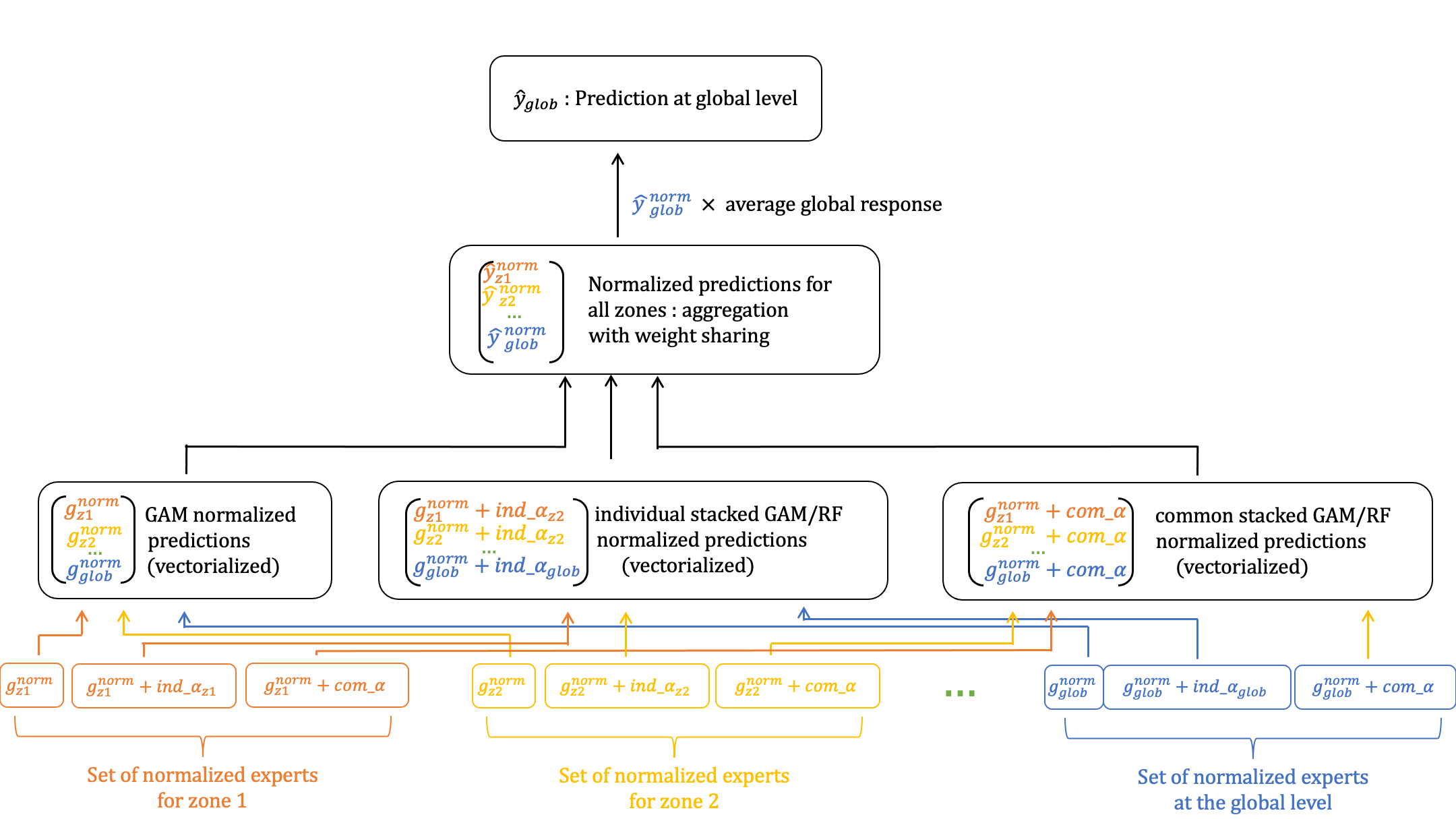}
                         }       
 \end{center}
  \caption{\label{fig:aggregation1} Fully disaggregated and Vectorial aggregations strategies.}
\end{figure}

\begin{figure}[H]
%\ContinuedFloat
  \begin{center}
    \subfloat[Unscaled hierarchical aggregation.]{
    \includegraphics[width=\textwidth]{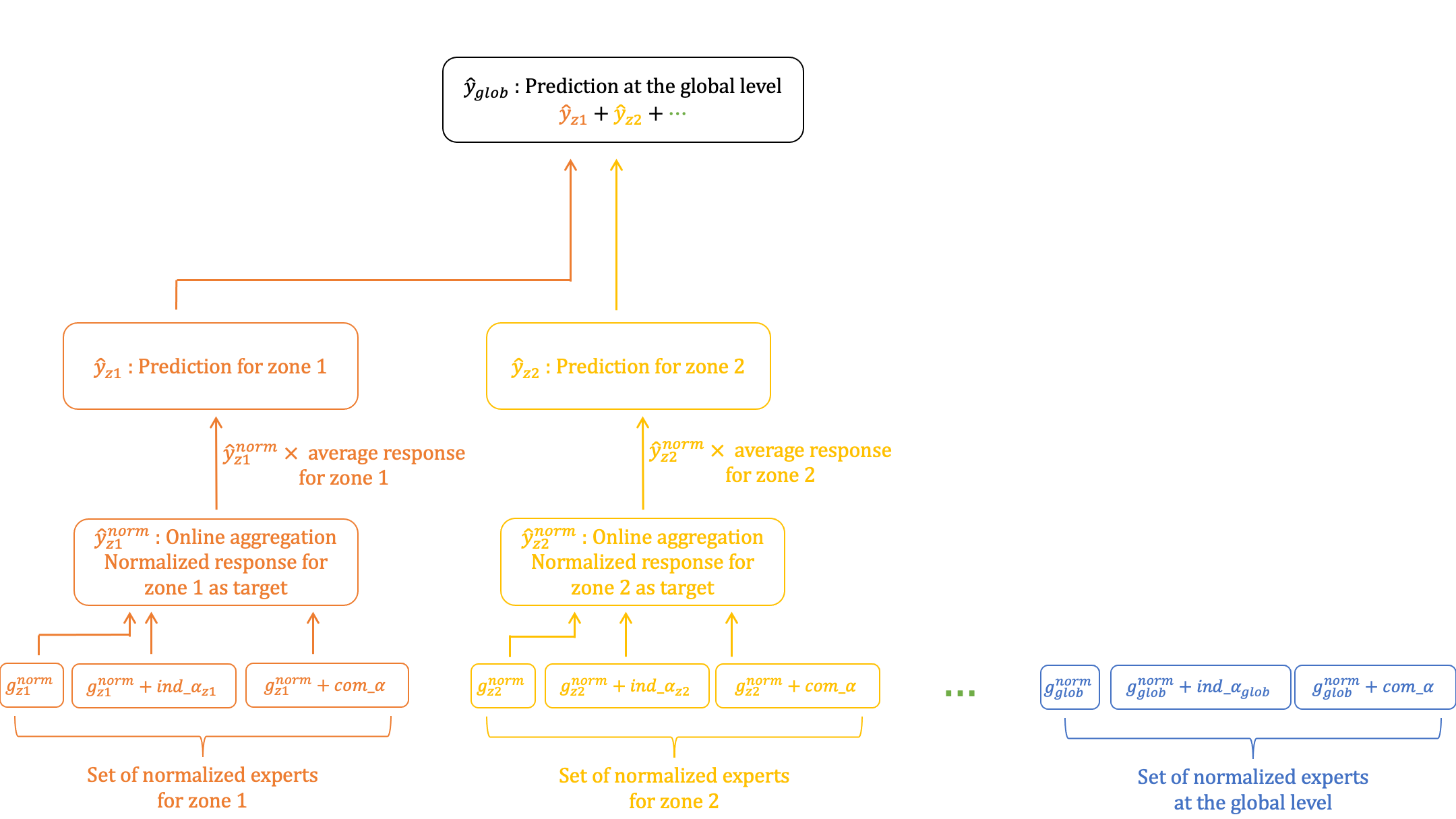}
                         }
                         
   \subfloat[Scaled hierarchical aggregation.]{
    \includegraphics[width=\textwidth]{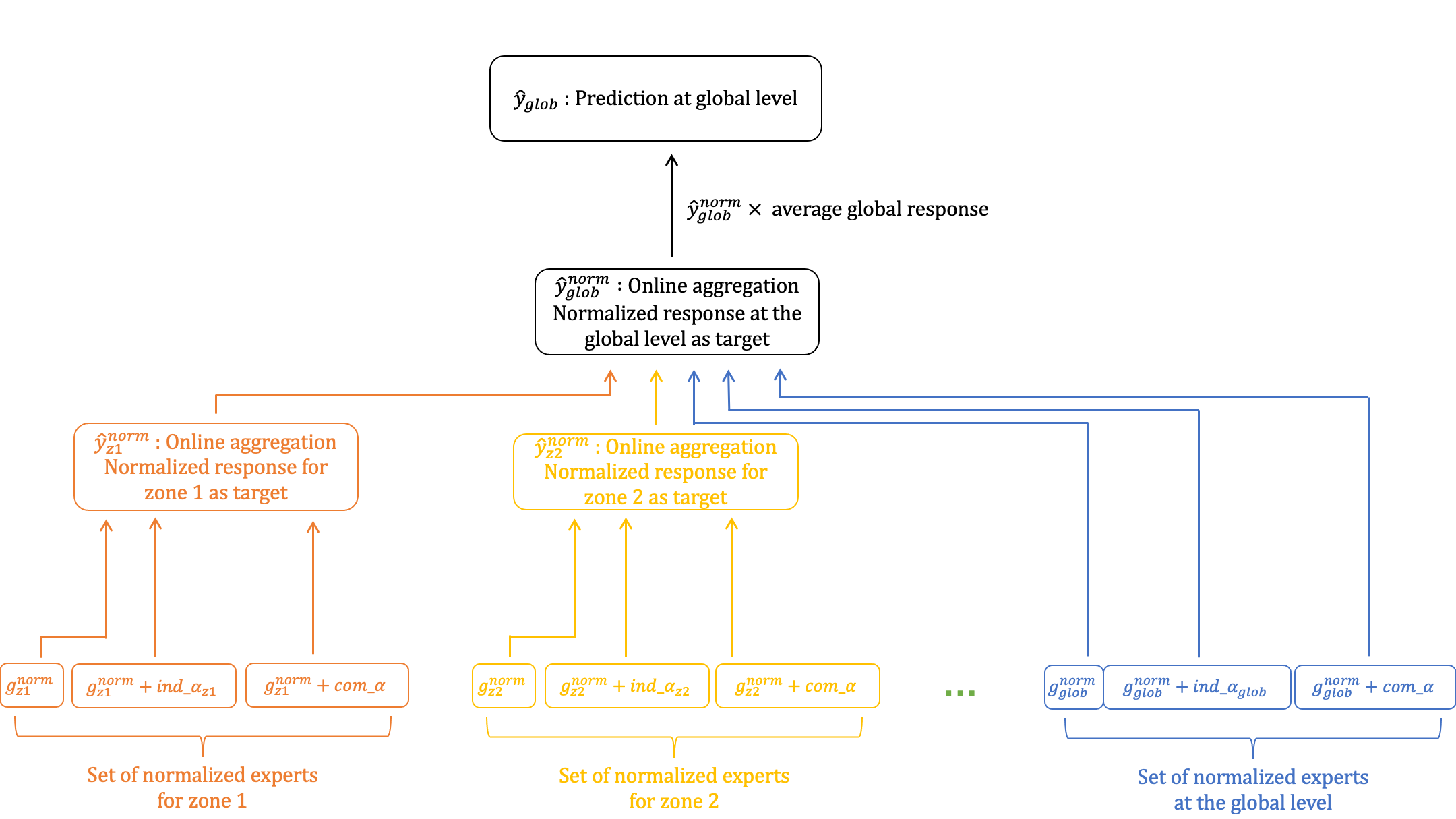}
                         }                    
 \end{center}
 \caption{\label{fig:aggregation2} Unscaled and scaled hierarchical Aggregations strategies.}
\end{figure}

The strategy described in Section \ref{subsec:experts} yields 11 experts for each one of K zones and for the global level: 1 GAM expert, 5 GAM-RF stacked experts trained zone by zone, and 5 GAM-RF stacked experts trained on the aggregated data. Thus, we obtain $11 \times K$ experts. To combine the predictions of these experts, we propose 4 aggregation strategies taking differently into account the hierarchical structure of the data. The algorithms are described below and illustrated in Figures \ref{fig:aggregation1} and \ref{fig:aggregation2}.

\begin{itemize}
    \item \textbf{Full disaggregated model}: we use the full set of $11(K+1)$ scaled forecasts as experts and the scaled global response $y^{norm}$ as our target variable. The prediction is then multiplied by the average value of the response at the global level.
    
    \item \textbf{Vectorial aggregation}: we illustrate the possibility to share weights between the zones and at the global level. We aim at predicting the time series of the (K+1)-dimensional vector corresponding to the scaled response in each zone and at the global level. To do so, we aggregate 11 vectorial, (K+1)-dimensional experts corresponding to the predictions of the GAMs and of the 10 stacked GAM-RF. The prediction corresponding to the global level is then multiplied by the average value of the response at the global level to forecast $y$.
    
    \item \textbf{Hierarchical aggregation, scaled predictions}: as a first step, we aggregate the 11 experts in each zone using the scaled response for a zone $z$, $y^{norm}_{z}$ as a target, and we obtain $K$ experts. Then, we aggregate these $K$ experts and the quantile experts at the global level, with the scaled global response $y^{norm}$ as our target variable. The prediction is then multiplied by the average value of the response at the global level.
    
    \item \textbf{Hierarchical aggregation, unscaled predictions}: we again aggregate the 11 experts in each zone, using the scaled response for the corresponding zone as a target, and we obtain $K$ experts predicting the normalized response at the zonal level. Then, we multiply their predictions by the average value of the response for the corresponding zone $y_{z}$, and sum these predictions in order to obtain a forecast of the global level $y$. We underline that this aggregation method is the only one enforcing coherency between the prediction at the zonal level and the prediction at the global level.
\end{itemize}

In Section \ref{sec:Covid}, we apply this methodology to adaptively predict electricity load at the national level in France during the first Covid lockdown using data at the regional level. We underline that our objective is the forecast at the national scale, and that forecasts at the regional scale are only used to improve this aggregated forecast. For this reason, we do not require that the forecasts be coherent (the sum of the forecasts at the regional level does not necessarily have to be equal to the forecast at the national level). In fact, the experiments presented in Section 5 indicate that aggregation methods that do not respect coherency (such as vectorial aggregation) can obtain more accurate results than methods that do (i.e. hierarchical aggregation with unscaled predictions).
\section{Transfer learning for forecasting aggregated smart meter data}
\label{sec:smartmeter}

In this section, we will illustrate the methodology using a dataset that is commonly used for the calibration of electricity consumption forecasting models. The dataset is made up of aggregate semi-hourly consumption data of the national load for the United Kingdom, and of observations of some meteorological and calendar variables. Our goal is to forecast electricity consumption at the national level from December 2009 to August 2010 (this period will be referred to as the test set). For this purpose, we assume that we have access to data at the national level covering the period from April 2005 to November 2009 (called hereafter the learning set) and data from smart meters for a smaller period (from April 2009 to August 2010). In this first example, we compare the performances of a GAM, a RF and a stacked GAM-RF trained using data at the national scale to the performances of a stacked GAM-RF using features learned from smart meter data. This allows us to highlight both the advantage of stacking GAM and RF and that of transferring the GAM features learned at the finer scale, and to decompose the contribution due to stacking GAM and RF and to using these new features.

\subsection{Data}

\subsubsection{National Data}
This dataset for UK national semi-hourly electricity consumption is
provided by the European Grid Standards Office (see
\url{https://www.nationalgrideso.com/balancing-data/data-finder-and-
explorer}) and covers the period between April 2005 and December 2010. We add as features the temperature data obtained from the NOAA (National Oceanic and Atmospheric Administration) \footnote{https://www.noaa.gov/} for the 10 largest cities in the UK: London, Birmingham, Glasgow, Sheffield, Bradford, Liverpool, Edinburgh, Manchester, and Bristol. We then compute a weighted average $T_t$ of the temperatures recorded in these 10 stations with weights proportional to the official population of each city, and we finally perform an exponential smoothing of this weighted average with the parameters 0.2, 0.05, and 0.01.

\subsubsection{Smart meters Data}
This data set corresponds to smart meters data at an individual scale in the UK. This dataset has been obtained from the Energy Demand Research Project (EDRP) launched by Ofgem on behalf of the UK Government in 2007 (see \cite{AECOM18}, \cite{schellong2011energy} and \footnote{\url{https://www.ofgem.gov.uk/gas/retail-market/metering/transition-smart-meters/energy-demand-research-project}}) where the power consumption of approximately 60,000 households was collected at half hourly intervals for about two years. We consider a subset of 1925 customers from April 2009 to August 2010 located in two regions of the UK: south-east (arround Brighton) and north-west (around Glasgow). %In addition to the electricity consumption, NUTS codes (Nomenclature of Territorial Units for Statistics) of level 4 are provided, along with the Acorn category value (integer between 1 and 6), to the type of heating fuel (gaz, electricity or dual) and to the contract type (Standard or Households containing an electricity meter with a time of use tarif) for each household. visibilities and humidities
We considered temperatures in each region, obtained from the NOAA. We add to this data set supplementary calendar covariates such as the time of year, day type, sunrise and sunset time along the year.

\subsection{Models and forecasting}

The fitting procedure used to forecast electricity consumption at the national level can be described as follows. 

%\medskip

We note  a trend in the time series of consumption over the period from April 2005 to August 2010. We estimate this trend in a very simple way by fitting to the series of observations a nonparametric Gaussian model $Y_t= \mu+s(t)+\varepsilon_t$, the trend $s(t)$ being represented in a cubic spline function base with a number of knots limited to three. In the following, we subtract this trend and aim at forecasting the national \textit{de-trended} consumption, which is then given by $Y_t^{c} =Y_t-\widehat{s}(t)-\hat{\mu}$.

%\begin{figure}[!h] 
%\centering
%\includegraphics[height=0.4\textheight]{TrendNat.pdf}
%\caption{The national electricity consumption over the observed period and its smoothed trend (in red)}
% \label{fig:trend} 
% \end{figure}
 
We apply the stacked GAM and RF methodology to predict national load consumption, using only data available at the national level. Note that this is a special case of the general transfer learning framework, with $\mathcal{D}_{\cT}=\mathcal{D}_{\cS}$, and where the final forecast is obtained using RF on the data enriched with the transfer of information performed using the design of new features. We begin by fitting a semi-parametric GAM on the learning set. The GAM is given by
\begin{equation}
	\begin{array}{lll}
			Y_t^{c} & = & \sum_{j=1}^{7}m_j \I_{{\rm DayType}_t=j}+m_8\I_{\rm{Holiday}_t=1}+m_9\I_{{\rm LongWeekEnd}_t=1}\\		
				 & + & g_1({\rm Instant}_t, {\rm Temp}_t)\ + \sum_{j=1}^{7} f_j({\rm Instant}_t)\I_{{\rm Daytype}_t=j} + s({\rm ToY}_t)\\
			    &+ & s({\rm Temp99}_t) + \varepsilon_t\\	
		\end{array}
		\label{national:GAM}
\end{equation}

where the variables are presented in Table \ref{tab:var_smart_meter}, and $\varepsilon_t$ is a centered Gaussian noise. Each univariate smooth component of the above GAM model is fitted using regression spline functions with 40 knots (50 knots for ${\rm \text{ToY}}$) and a tensor basis of spline functions for the interaction between time and temperature with 20 and 10 knots, respectively.

\begin{table}
\begin{center}
\begin{tabular}{ l|l }
$Y_t^{c}$& de-trended electricity load \\ \hline
${\rm Temp}_t$ & weighted temperature\\ \hline
${\rm Temp99}_t$ & weighted and exponentially
smoothed temperature\\ \hline
 ${\rm Instant}_t$ & instant of the day\\ \hline
 ${\rm DayType}_t$ & day of the week\\ \hline
${\rm Holiday}_t$ & binary variable
indicating public holiday days\\ \hline
${\rm LongWeekEnd}_t$ & binary
variable indicating the presence of a long weekend\\ \hline
${\rm ToY}_t$ & time of year\\
\end{tabular}
\end{center}
\caption{\label{tab:var_smart_meter} Variables used in the models \ref{national:GAM} and \ref{local:GAM_total}}
\end{table}

Then, once the fit~(\ref{national:GAM}) has been performed, we extract the estimated effects $g_1$, and $f_j$ of the features, and add them to the set of initial covariates. This enriched dataset will then be used to train the RF. Note that the initial number of covariates in the database taken into account in the Equation~(\ref{national:GAM}) is 7, and the number of additive components extracted by the GAM methodology is 15. We thus find ourselves, after transfer, with a sample having a number of observed covariates equal to 22. We apply the GAM-RF stacking methodology to fit a nonparametric regression called {\tt GAM.RF.nat} on the training sample. Finally, we evaluate the prediction of this stacked GAM and RF on the test sample, and compare it to the predictions obtained with the GAM~(\ref{national:GAM}) model. We also compare this model to a standard regression model by random forests, denoted hereafter {\tt RF.nat}.

%\smallskip
In a second time, we apply the stacked GAM and RF methods to transfer information from the smart meters data. We begin by computing the total consumption of the customers on the smart meter data set, and we fit a GAM model to forecast this total. Using the same methodology as for the national data, we obtain the  model presented in Equation \eqref{local:GAM_total}. We chose to use a simpler model than the national one, because the dataset used to train it is smaller. 

\begin{equation}
	\begin{array}{lll}
			y_t & = & \sum_{j=1}^{7}m_j \I_{DayType_t=j}\\		
				 & + & g_1({\rm Instant}_t, T_t)\ + \sum_{j=1}^{7} f_j({\rm Instant}_t)\I_{DayType_t=j} + s(\text{ToY}_t)\\
			    &+ & \varepsilon_t\\	
		\end{array}
		\label{local:GAM_total}
\end{equation}
We then extract the 10 non-linear features of this model as supplementary covariates and add them to the dataset consisting of all original covariates and of the effect extracted from the GAM at the national level. Finally, we use these covariates to train the stacked GAM-RF, and obtain a model called {\tt GAM.RF.local}. We underlined that the GAM \eqref{local:GAM_total} used to model the aggregated smart-meter load is fitted on the small dataset consisting of smart-meter data. However, the effects $f_k$ extracted from this model are then used to create new features for each entry of the large national dataset. Thus, the models RF.nat, GAM.nat, GAM.RF.nat, and GAM.RF.local are trained on the same number of observations from the national dataset.

\begin{table}[!htbp] 
\centering 
  \label{ref:tab1} 
\begin{tabular}{@{\extracolsep{5pt}} ccccc} 
\\[-1.8ex]\hline 
\hline \\[-1.8ex] 
 & GAM.nat & RF.nat & GAM.RF.nat & GAM.RF.local\\ 
\hline \\[-1.8ex] 
RMSE & $1 409$ MW & $1 339$ MW & $1 214 $ MW & $1 193$ MW \\ 
MAPE & $2.670$ & $2.560$ & $2.360$ & $2.310$\\ 
Nb of covariates & $7$ & $7$ & $22$ & $32$\\ 
\hline \\[-1.8ex] 
\end{tabular} \caption{\label{tab:res_smartmeter} Errors of prediction for the learners {\tt GAM.nat}, {\tt RF.nat}, {\tt GAM.RF.nat}, and {\tt GAM.RF.local}}   

\end{table} 

An importance by permutation analysis of the variables used in the stacked RF after learning by transfer retains as most important for {\tt GAM.RF.nat} the instant of the day, followed by three terms from the national GAM modeling. For the model {\tt GAM.RF.local}, the most important variables are the instant of the day, followed by two terms from the national GAM, and by one term from the local GAM.

By analyzing the mean absolute percentage errors (MAPE) and the root mean squared errors (RMSE) of the different methods, presented in Table \ref{tab:res_smartmeter}, we see that for the British national data set, the RF are more efficient than the adopted reference model GAM. Interestingly, the stacked GAM and RF trained using only national data {\tt GAM.RF.nat} outperforms these two models. This indicates that the stacking of GAM and RF allows to obtain the best of both worlds: the RF is able to correct effects or interactions of variables (such as the instant of the day) that are not well captured by the GAM, while being robust to the large number of covariates taken as input (up to 28). Finally, the best model both in terms of MAPE and RMSE is obtained by stacking GAM and RF using effects learned from both national and smart meter data. These results underscore the value of leveraging available data at a finer scale, even when  no hierarchical constraints are implemented in the algorithm.

%%%%%%%%%%%%%%%%%%%%%%%%%%%%%%%%%%%%%%%%%%%%%%%%%%%%
%%%%%%%%%%%%%%%%% APPLICATION: French electricity data %%%%%%%%%%%%%%%%% 
%%%%%%%%%%%%%%%%%%%%%%%%%%%%%%%%%%%%%%%%%%%%%%%%%%%%
%

\section{Electricity load forecasting during the first Covid-19 lockdown}
\label{sec:Covid}
In this Section, we apply our methodology to short-term electricity load forecasting during the Covid-19 lockdown and post-lockdown period in France, at a resolution of half an hour and at the national level. To do so, we leverage information available at the regional level. Electricity consumption has been significantly affected by the measures taken by the government to cope with the epidemic: closures of non-essential businesses, as well as stay-at-home directives, have led to a decrease in electricity consumption of about 10\%, as well as to changes in its daily and weekly patterns (see \cite{obst2021adaptive} for a description of the impact of these measures on french electricity consumption). Common models trained on historical data, which rely on calendar and weather data, fail to account for these significant changes. Similarly, transfer learning methods relying on data present at a finer (e.g., regional) scale, if trained on data with different distribution than that of the target, will make poor predictions, especially if the relationship between local and global variables changes over time. Thus, these models, trained on data from the pre-pandemic period, make relatively large prediction errors on the period following the start of the lockdown. To ensure adaptativity of our models, we combine the stacked GAM-RF methodology presented in Section \ref{subsec:stackedGAMRF} with the online aggregation of quantile experts presented in Section \ref{subsubsec:aggregation}.

Transfer learning proves to be essential to address the problem of electricity load forecasting during the Covid-19  pandemic. Indeed, as the data for this period is scarce, especially since we want to make predictions from its very beginning, it is crucial to use information from the pre-pandemic period to predict power consumption during the pandemic period. To do so, we use the methods presented above to transfer information from the large data set corresponding to historical electricity consumption during the pre-pandemic period (hereafter called the source period) to improve predictions during the pandemic period (called the target period). This can be done using again stacked GAM and RF: this transfer learning algorithm allows us to rely on a GAM trained on a large set of observations of historical electricity consumption, coming from the source distribution, while correcting its error on the target using RF based on scarce observations.

On the other hand, because of the important changes in electricity consumption consecutive to the lockdown, we expect that the relationship between effects learned on regional data and national load will also change. Indeed, our studies reveal that containment measures induce changes in electricity consumption at the regional level, which however differ according to the region considered. To make use of electricity consumption data available at the regional level, we must remain adaptative to changes in the distribution of both national and regional data. This is achieved by using online aggregation of experts, which allows us to combine forecasts at the regional and national levels in an adaptative fashion. We choose to forecast electricity consumption separately region by region using stacked GAM and RF, and then combine the forecasts of these regional models in order to predict national electricity consumption in a hierarchical fashion. In doing so, the hierarchical model captures regional phenomena that are not apparent at a more aggregated scale and leverages this information to improve predictions at the national level. Thus, our methods allow for transferring knowledge both at a temporal level (data from the pre-pandemic period are used to improve forecasts during the pandemic period), and at a hierarchical level (regional predictions are used to produce forecasts at the national level). 

The rest of the section is organized as follows. In Section \ref{subsec:data}, we present the data used to design and evaluate our models. In Section \ref{subsec:modelsCovid} we present the models used for forecasting electricity consumption at the national and regional levels. The results of our study are presented in Section \ref{subsec:resultsCovid}: first, we compare the performances of different approaches, then we present a more detailed analysis of the stacked GAM and RF, and of the online aggregation of experts.

\subsection{Data}\label{subsec:data}

The data are from the french TSO (Transmission  System Operator)  \href{https://opendata.reseaux-energies.fr/}{RTE}. It consists of electricity consumption (in MW) at a half-hourly temporal resolution at the French national level (``Load") and for the 12 metropolitan administrative regions (it does not include Corsica): Nouvelle Aquitaine, Auvergne Rhônes-Alpes, Bourgogne-Franche-Comté, Occitanie, Hauts-de-France, Normandie, Bretagne, Centre-Val de Loire,  Île-de-France, Pays de la Loire, Provence-Alpes-Côte d'Azur, Grand Est. Our goal is to forecast the French national consumption, exploiting the regional loads information. For all the load consumption data, we compute the lags for one day and one week and denote it with the subscripts``.48" and ``.336".

Our models use the temperature and weighted temperature as explanatory variables. These variables were collected on the website of the French weather forecaster \href{https://donneespubliques.meteofrance.fr/}{Météo France}.
For each region, we compute the weighted mean of meteorological stations where the weights are proportional to $\exp(-dist)$ where $dist$ is the distance of the station to the barycenter of each region. Note that we use the observed temperatures instead of their predicted values in our forecast. In doing so, we cancel out the errors caused by the uncertainty of a particular weather forecast, which allows for a more precise comparison of the different models. Moreover, this choice allows us to only use open data, so as to ensure reproducibility of our results.

Our models also rely on variables characterizing the impact of the restrictions implemented to fight the epidemic. The first of these variables is the Oxford Covid-19 Government Response Tracker. This index, freely available at \url{https://www.bsg.ox.ac.uk/research/research-projects/covid-19-government-response-tracker}, aggregates indicators characterizing the measures taken by governments to mitigate the epidemic in terms of containment, health, and economic support. It is available at the national level. The methodology used to calculate the index and the measures on which it is based are known a few days in advance, so we assume that it is known for the day we wish to forecast. The remaining variables used to characterize the impact of lockdown measures are Google Mobility Indices. These indices are provided by Google, and obtained by aggregating geolocalisation data. They characterize the changes in the frequentation of categorized places (residential, workplaces, transports, parks, grocery and pharmacy, retail and recreation). The data are freely available at \url{https://www.google.com/covid19/mobility/}, albeit with a little less than a week delay. Therefore, we considered lagged versions of these indicators in our prediction. The government response and mobility indices are available respectively from January and February 2020 onwards. Therefore, we do not use them as covariates in the source model, but only in the target model.

Hereafter, we call source models the models trained on historical data, collected between the beginning of 2012 to the end of August 2019. We evaluate their performances on data with the same distribution, during the pre-lockdown period ranging from September 2019 to March 15th 2020, and compare it to their performance on data from the target distribution, ranging from March 16th to September 17th. By contrast, the models specific to the lockdown and post-lockdown period, henceforth called target models, are retrained every day during this target period, so as to leverage all observations available. Note that while the first french lockdown officially started on March 17th, we consider March 16th as the first day of the target distribution, as the electricity consumption pattern had already changed by that day.

\subsection{Models}\label{subsec:modelsCovid}
%We proceed in three steps: 
%\begin{enumerate}
%  \item First, we fit a GAM (described below) on the source data. We use it to produce predictions of the load on the target data, and compute the residuals $\widehat{\varepsilon}_t$ during the pandemic period. Moreover, we extract the estimated GAM effects of the covariates for this period.
%  \item Then, each day from the first day of containment, we train RF (described below) on the available data for the period of the pandemic. These RF are trained to predict the residuals $\widehat{\varepsilon}_t$ of the historical GAM on the target data, and take as input the estimated GAM effects, the original covariates and the mobility indices.
%  \item Finally, we combine the predictions of 
%\end{enumerate}

%Using RF to correct the errors of the GAM during the pandemic period allows us to obtain an adaptative model able to produce predictions from the very beginning of the target period. Note that the corrections of the RF remain small compared to the predictions of the GAM: the first order of the prediction is given by the source model, trained on the large set of historical data, while the corrections learned on the scarce observations from the target dataset only provide a second order correction.

\subsubsection{Generalized additive models for the pre-pandemic period}\label{subsec:GAMCovid}

We use GAM to predict the electricity load under normal circumstances. We fit one model for each region of mainland France, as well as one at the national level, and obtain thus 13 models. To take into account the daily patterns of electricity consumption, each model is composed of 48 GAM fitted independently and forecasting the electricity load at a given instant of the day. Thus, the 624 time-series corresponding to the 48 half-hours for the 12 regions and the national level are treated independently. In order to compare the predictions, terms, and errors of the models, regional and national electricity loads are normalized, that is, they are divided by their average value for the region and the half-hour considered. GAM are then fitted to predict this normalized load. In the following, we denote respectively by $y$ and $y^{norm}$ the load and the \textit{normalized} load.

The model used to predict the electrical load for a zone $z$ at a time $t$ corresponding to the $h$-th half-hour of the day is the following:
\begin{eqnarray}\label{eq:gam_covid}
y_{z,t}^{norm} &=& \underset{i = 1}{\overset{7}{\sum}} \underset{j = 0}{\overset{1}{\sum}} \alpha_{i,j}^{(z,h)}\mathds{1}_{\text{DayType}_t = i}\mathds{1}_{\text{DLS}_t = j} + \underset{i = 1}{\overset{7}{\sum}} \beta_{i}^{(z,h)}\text{Load.48}_{z,t}\mathds{1}_{DayType = i} \\
&& + \gamma^{(z,h)}\text{Load.336}_{z,t} + f_1^{(z,h)}(t) + f_2^{(z,h)}(\text{ToY}_t) + f_3^{(z,h)}(t, \text{Temp}_{z,t}) \\
&& + f_4^{(z,h)}(\text{Temp95}_{z,t}) + f_5^{(z,h)}(\text{Temp99}_{z,t}) + f_6^{(z,h)}(\text{TempMin99}_{z,t}, \text{TempMin99}_{z,t}) + \varepsilon_{z,t}
\end{eqnarray}

\begin{table}
\begin{center}
\begin{tabular}{ l|l }
 $y_{z,t}^{norm}$& normalized electricity load for the zone $z$ \\ \hline
 Daytype$_t$ & categorical variable indicating the day of the week \\ \hline
 DLS$_t$ & binary variable indicating whether $t$ is in summer hour or winter hour\\ \hline
${\rm ToY}_t$ & time of year\\ \hline
Temp$_{z,t}$ & temperature in the zone $z$\\ \hline
Temp95$_{z,t}$ & weighted and exponentially
smoothed temperature of smoothing factor 0.95\\ \hline
Temp99$_{z,t}$ & weighted and exponentially
smoothed temperature of smoothing factor 0.99\\ \hline
TempMin99$_{z,t}$ & minimal value over the day of Temp99$_{z,t}$\\ \hline
TempMax99$_{z,t}$ & maximal value over the day of Temp99$_{z,t}$\\ \hline
Load.48$_{z,t}$ & normalized load of the day before in the zone $z$\\ \hline
Load.336$_{z,t}$ & normalized load of the week before in the zone $z$
\end{tabular}
\end{center}
\caption{\label{tab:var_covid} Variables at time $t$ used in model \ref{eq:gam_covid}.}
\end{table}

where $\varepsilon_{z,t}$ is gaussian white noise, and the variables are presented in Table \ref{tab:var_covid}. Each univariate smooth component of the above GAM model is fitted using regression spline functions with respectively 20 knots for ToY, 10 knots for Temp95 and Temp99, 5 knots for Date,  and a tensor basis of spline functions for the interaction between time and temperature with 3 and 5 knots, respectively.

\subsubsection{Quantile GAM-RF experts aggregation}\label{subsec:stackedCovid}
We design experts by stacking GAM and RF, following the methodology described in Section \ref{sec methodology}. The RF are trained in a streaming fashion on the target data (pandemic and post-pandemic period). Building on the results of Section \ref{sec:smartmeter}, we choose to take as input for these RF the usual covariates, but also the GAM effects learned on the source dataset. Interestingly, preliminary results postponed to \ref{app:selectionVariables} indicate that while the RF inputs are high-dimensional, variable selection only marginally affects the performance of our model. The RF appear to be robust against the high dimension of the features, even in the early days of lockdown, where few observations are available.

Using RF to correct the errors of the GAM during the pandemic period allows us to obtain an adaptative model able to produce predictions from the very beginning of the target period. Note that the corrections of the RF remain small compared to the predictions of the GAM: the first order of the prediction is given by the source model, trained on the large set of historical data, while the corrections learned on the scarce observations from the target dataset only provide a second order correction.

For each of the 12 regions and at the national level, we obtain 
11 experts, corresponding to the GAM experts, the 5 GAM-RF quantiles experts trained on the residuals of the zone, and the 5 GAM-RF quantiles experts trained on the aggregated residuals. Then, we compare 4 aggregation techniques (full disaggregated model, vectorial aggregation, hierarchical aggregation of scaled predictions, and hierarchical aggregation of unscaled predictions).

\subsection{Results}
\label{subsec:resultsCovid}
In this Section, we compare the methods presented above. First time, we compare their performances in terms of MAPE and RMSE in Section \ref{subsec:perfCovid}. Then, we lead an importance by permutation analysis of the RF, which is presented  in Section \ref{subsec:stackedCovid}. Finally, we analyze and compare the different aggregation methods in Section \ref{subsec:aggregationCovid}.

\subsubsection{Performances}\label{subsec:perfCovid}
In Table \ref{table:MAPE}, we compare the MAPE and the RMSE of the 4 methods, of the GAM at the national level, and of the stacked individual RF predicting the median of the residuals at the national level.

We split the test period into 3 sub-periods. In the pre-pandemic period between September 1st 2019 and March 15th 2020, only the GAM predictions are available. During this period, vectorial aggregation makes little sense since there is only one type of expert. The lockdown period ranges between March 16th and May 11th: during this period, training data are very scarce, and models must quickly adapt to a dramatic change in electricity consumption patterns. The post-lockdown period, from May 12th to September 17th, corresponds to a new change in load pattern, due to a relative rebound in activity, to which the models must adapt.

{
\small
\begin{table}[!h]
\begin{tabular}{c|c|c|c}
    \small{Model} & \small{2019/09/01-2020/03/15}&\small{2020/03/16-2020/05/11}&\small{2020/05/12-2020/09/17} \\
    \hline
    \small{GAM} & 1.36 \%, 1030 MW & 4.82 \%, 2838 MW & 1.84 \%, 1045 MW\\
    \hline
    \small{Individual stacked GAM-RF} & Non applicable & 2.41 \%, 1813 MW & 1.03 \%, 592 MW \\
    \hline
    \small{Full disaggregated} & 1.20 \%, 910 MW & 2.26 \%, 1716 MW & 1.09 \%, 609 MW \\
    \hline
    \small{Hierarchical aggregation} & 1.14 \%, 861 MW  & 2.21 \%, 1648 MW & 1.07 \%,  609 MW\\
    \small{scaled} &  & & \\
    \hline
    \small{Hierarchical aggregation} & 1.20 \%, 907 & 2.08 \%, 1553 MW & 1.02 \%,   593 MW\\
    \small{unscaled} &  & & \\
    \hline
    \small{Vectorial aggregation} &  Non applicable & 2.56 \%, 1885 MW & 0.91 \%,  521 MW\\
\end{tabular}
\caption{Mean Absolute Percentage Error and Root of the Mean Squared Error of the stacked GAM and RF models.}
\label{table:MAPE}
\end{table}
}

GAM-RF stacking improves GAM significantly. Using a stacked RF predicting the median of the GAM residuals at the national level is enough to decrease its MAPE during and after the lockdown by respectively 50 and 45\%. All hierarchical aggregation strategies, except for vector aggregation, outperform GAM and GAM-RF during the lockdown period: these results indicate that online aggregation is an efficient way to take into account information available at a finer scale. Our analysis in \ref{app:covid} shows that regional GAM have in average errors much larger than that of the GAM at the national level, due to larger fluctuations present at the finer scale. Interestingly, aggregating these low-accuracy models allows to obtain better performances than that of the GAM at the national level, even in the pre-pandemic period. This confirms the interest of aggregating quantile GAM-RF experts to track changes in the data. While vectorial aggregation performs rather poorly compared to other aggregation strategies during the lockdown, it improves over all other models after the end of the lockdown.

In \ref{app:covid}, we analyze the stacked GAM and RF. We plot the evolution of the importance of the variables across time. Our results show that variables important for predicting one quantile tend to be important for predicting the other quantiles. Moreover, the effects of the GAM are among the most important covariates for predicting the GAM residuals. Using these effects as covariates allows to transfer information on the impact of weather and calendar variables learned on the large dataset of pre-pandemic observations. We also note that as time passes and the size of the training set for the RF increases, relevant variables such as the Government Response Tracker, or relative occupation of some places of interest become more important for the prediction. Conversely, spurious variables are discarded as unimportant. Interestingly, the common RF trained on residuals across all regions detects these relevant variables more quickly than the individual RF trained solely on residuals at the national level. This highlights the interest of aggregating the data across zones and scales in this sparse data context.

The fact that scaled and unscaled hierarchical aggregation obtain similar performances is somewhat counterintuitive, given that in the scaled model the aggregation must learn the contribution of the different regions to the national consumption. To analyze this phenomenon, we investigate in \ref{app:covid} the relative weight given to the experts corresponding to the different regions. We find that the weights in the unscaled hierarchical aggregation do not correspond to the proportions of electricity consumed by the regions, and that they typically exhibit much more flexibility than the true weights. The fact that the scaled hierarchical aggregation outperforms its unscaled counterpart both in the pre-pandemic and in the post-lockdown period suggests that the flexibility provided by the second layer of aggregation used in the scaled model compensates for the lack of knowledge of the relative contribution of the different regions.

We also study the weights given by the experts in the vectorial aggregation, and note that it gives a predominant weight to the GAM and median staked RF experts, which appears as the most relevant experts across all regions; however, these weights are highly unstable during the beginning of the lockdown. The performance of the vectorial aggregation during this period is worst than that of all other aggregation models, and than that of the stacked RF predicting the median of residuals. This behavior mirrors the fact that the impact of the pandemic strongly differs from one region to another, as is shown in \ref{app:covid}. On the other hand, vectorial aggregation achieves the best performance during the post-lockdown period and appears as a promising approach to predicting consumption under normal circumstances.

\section{Conclusions and Future Work}
\label{sec:Conclusion}

We propose new transfer learning methods designed for forecasting time series  observed  at different hierarchical scales. We present two different settings and illustrate them with two different usecases:

\begin{enumerate}
    \item To transfer information from finer scale (an aggregate of smart meters)  to wider scale (national) data when the distribution of the data is stable with time, we propose to stack  features from GAM obtained at these two scales into random forests. 
    \item To transfer information from local to global data when the distribution of the data is changing with time we propose hierarchical online aggregation of experts where the experts are generated at a finer scale (regional level) using quantile stacked random forest. 
\end{enumerate}

We demonstrate the interest of our  proposed approach in  both  cases. In both cases transfer learning by RF stacking at a single scale improve significantly the forecasting performance of single GAM or RF model: 14\% of improvement over GAM and 9\% over the RF for case 1, 38\% over GAM for case 2. It supports our original intuition that stacked RF gather both the ability of GAM to extrapolate and RF to model automatically interaction between covariates.

Regarding multi-scale transfer performances, we also obtained convincing results. In case 1, we improved the day-ahead forecasting performance of the wider scale stacked GAM-RF of about 1.5\% with our multi-scale transfer algorithm. For case 2, the best hierarchical aggregation algorithm improves about 10\% the stacked GAM-RF at a wider scale. Our relatively simple strategy of re-scaling plus aggregation behaves well in this bi-level hierarchy. We also saw that introducing strong constrains in the aggregation weight (vectorial aggregation) can be an interesting transfer strategy when the experts behave similarly at the different scales of the hierarchy. This is true during the post-covid period but not during the hard lockdown in France of March-April 2020 (we suspect that the effect of COVID on the electricity load impacts the different regions in a desynchronized way). 

% Variable selection in the stacked RF didn't show any improvement  here but we believe that in high dimensional setting there is some opportunity to introduce an automatic selection procedure such as lasso selection (we tested this reinforce RF procedure on simulated data and obtained interesting results). 

The main learner used in the paper for the final forecasting is based on stacked GAM-RF. We could have chosen other machine learning methods such as tree-based gradient boosting or neural networks, which can be tested in future  work. Our experiments revealed that automatic variable selection when forecasting didn't show any improvement. However, in a high dimensional setting with a large number of features generated when learning the source, we believe that a possible approach that is worth exploring, is to use for forecasting a regression-reinforced random forest (RFRF) approach that may have better prediction performance than RFs. The idea behind RFRF is to borrow the strength of penalized parametric regression to improve RF. For example, for RFRFs, we may run a SCAD (or LASSO) (see \cite{FanLi2001}) based selection before RF, then construct a RF on the residuals from the SCAD (or LASSO) penalized fit. Preliminary simulation results show that RFRFs can capitalize on the strength of both parametric and nonparametric methods and may give reliable predictions in high-dimensional extrapolation problems such as those encountered in transfer learning.

In our first usecase, we did not investigate the clustering of smart meter data to generate diverse GAM features, but this is clearly a possible improvement. Introducing hierarchical constrains in the weights, as proposed in \cite{bregere2021online}, is an other potential perspective for our second usecase. Finding the good warping of weights constraints for vectorial aggregation could also be a way to improve the performance of this method on desynchronized data.

\section*{Acknowledgements}
The authors would like to thank the associate editor and the anonymous reviewers for their valuable suggestions and comments.
%Vectorial aggergation fails to account for the difference in the errors for the regional GAM: use stacking with covariates?

%\section*{Acknowledgement}

%The authors would like to thank the editors, associate editors and the reviewers for their very helpful comments leading to major improvements of the  paper. The authors acknowledge the support of the French Agence Nationale de la Recherche (ANR) under reference ANR-20-CE40-0025-01 (T-REX project).
\newpage

\bibliography{biblio_transfer.bib}

\begin{thebibliography}{38}
\expandafter\ifx\csname natexlab\endcsname\relax\def\natexlab#1{#1}\fi
\providecommand{\url}[1]{\texttt{#1}}
\providecommand{\href}[2]{#2}
\providecommand{\path}[1]{#1}
\providecommand{\DOIprefix}{doi:}
\providecommand{\ArXivprefix}{arXiv:}
\providecommand{\URLprefix}{URL: }
\providecommand{\Pubmedprefix}{pmid:}
\providecommand{\doi}[1]{\href{http://dx.doi.org/#1}{\path{#1}}}
\providecommand{\Pubmed}[1]{\href{pmid:#1}{\path{#1}}}
\providecommand{\bibinfo}[2]{#2}
\ifx\xfnm\relax \def\xfnm[#1]{\unskip,\space#1}\fi
%Type = Techreport
\bibitem[{AECOM(2018)}]{AECOM18}
\bibinfo{author}{AECOM} (\bibinfo{year}{2018}).
\newblock {\it \bibinfo{title}{Energy Demand Research Project: Early Smart
  Meter Trials, 2007-2010}\/}.
\newblock \bibinfo{type}{Technical Report} UK Data Service.
%Type = Article
\bibitem[{Amato et~al.(2017)Amato, Antoniadis, De~Feis \&
  Goude}]{amato2017estimation}
\bibinfo{author}{Amato, U.}, \bibinfo{author}{Antoniadis, A.},
  \bibinfo{author}{De~Feis, I.}, \& \bibinfo{author}{Goude, Y.}
  (\bibinfo{year}{2017}).
\newblock \bibinfo{title}{Estimation and group variable selection for additive
  partial linear models with wavelets and splines}.
\newblock {\it \bibinfo{journal}{South African Statistical Journal}\/},  {\it
  \bibinfo{volume}{51}\/}, \bibinfo{pages}{235--272}.
%Type = Article
\bibitem[{Amato et~al.(2021)Amato, Antoniadis, De~Feis, Goude \&
  Lagache}]{amato2021forecasting}
\bibinfo{author}{Amato, U.}, \bibinfo{author}{Antoniadis, A.},
  \bibinfo{author}{De~Feis, I.}, \bibinfo{author}{Goude, Y.}, \&
  \bibinfo{author}{Lagache, A.} (\bibinfo{year}{2021}).
\newblock \bibinfo{title}{Forecasting high resolution electricity demand data
  with additive models including smooth and jagged components}.
\newblock {\it \bibinfo{journal}{International Journal of Forecasting}\/},
  {\it \bibinfo{volume}{37}\/}, \bibinfo{pages}{171--185}.
%Type = Article
\bibitem[{Anderer \& Li(2022)}]{anderer2022hierarchical}
\bibinfo{author}{Anderer, M.}, \& \bibinfo{author}{Li, F.}
  (\bibinfo{year}{2022}).
\newblock \bibinfo{title}{Hierarchical forecasting with a top-down alignment of
  independent-level forecasts}.
\newblock {\it \bibinfo{journal}{International Journal of Forecasting}\/}, .
%Type = Techreport
\bibitem[{Balestriero et~al.(2021)Balestriero, Pesenti \&
  LeCun}]{balestriero2021learning}
\bibinfo{author}{Balestriero, R.}, \bibinfo{author}{Pesenti, J.}, \&
  \bibinfo{author}{LeCun, Y.} (\bibinfo{year}{2021}).
\newblock {\it \bibinfo{title}{Learning in High Dimension Always Amounts to
  Extrapolation}\/}.
\newblock \bibinfo{type}{Technical Report} arXiv:2110.09485.
%Type = Article
\bibitem[{Breiman(2001)}]{breiman2001random}
\bibinfo{author}{Breiman, L.} (\bibinfo{year}{2001}).
\newblock \bibinfo{title}{Random forests}.
\newblock {\it \bibinfo{journal}{Machine learning}\/},  {\it
  \bibinfo{volume}{45}\/}, \bibinfo{pages}{5--32}.
%Type = Book
\bibitem[{Breiman et~al.(1984)Breiman, Friedman, Olshen \&
  Stone}]{breiman1984cart}
\bibinfo{author}{Breiman, L.}, \bibinfo{author}{Friedman, J.},
  \bibinfo{author}{Olshen, R.}, \& \bibinfo{author}{Stone, C.}
  (\bibinfo{year}{1984}).
\newblock {\it \bibinfo{title}{Classification and regression trees}\/}.
\newblock \bibinfo{publisher}{Chapman \& Hall/CRC}.
%Type = Article
\bibitem[{Brégère \& Huard(2021)}]{bregere2021online}
\bibinfo{author}{Brégère, M.}, \& \bibinfo{author}{Huard, M.}
  (\bibinfo{year}{2021}).
\newblock \bibinfo{title}{Online hierarchical forecasting for power consumption
  data}.
\newblock {\it \bibinfo{journal}{International Journal of Forecasting}\/},
  {\it \bibinfo{volume}{to appear}\/}.
%Type = Article
\bibitem[{Capezza et~al.(2021)Capezza, Palumbo, Goude, Wood \&
  Fasiolo}]{capezza2021additive}
\bibinfo{author}{Capezza, C.}, \bibinfo{author}{Palumbo, B.},
  \bibinfo{author}{Goude, Y.}, \bibinfo{author}{Wood, S.~N.}, \&
  \bibinfo{author}{Fasiolo, M.} (\bibinfo{year}{2021}).
\newblock \bibinfo{title}{Additive stacking for disaggregate electricity demand
  forecasting}.
\newblock {\it \bibinfo{journal}{The Annals of Applied Statistics}\/},  {\it
  \bibinfo{volume}{15}\/}, \bibinfo{pages}{727--746}.
%Type = Book
\bibitem[{Cesa-Bianchi \& Lugosi(2006)}]{Cesa-Bianchi:2006}
\bibinfo{author}{Cesa-Bianchi, N.}, \& \bibinfo{author}{Lugosi, G.}
  (\bibinfo{year}{2006}).
\newblock {\it \bibinfo{title}{Prediction, Learning, and Games}\/}.
\newblock \bibinfo{address}{New York, NY, USA}: \bibinfo{publisher}{Cambridge
  University Press}.
%Type = Article
\bibitem[{Dong et~al.(2021)Dong, Zhang, Wang \& Zhou}]{dong2021wind}
\bibinfo{author}{Dong, Y.}, \bibinfo{author}{Zhang, H.}, \bibinfo{author}{Wang,
  C.}, \& \bibinfo{author}{Zhou, X.} (\bibinfo{year}{2021}).
\newblock \bibinfo{title}{Wind power forecasting based on stacking ensemble
  model, decomposition and intelligent optimization algorithm}.
\newblock {\it \bibinfo{journal}{Neurocomputing}\/},  {\it
  \bibinfo{volume}{462}\/}, \bibinfo{pages}{169--184}.
%Type = Article
\bibitem[{Fan \& Li(2001)}]{FanLi2001}
\bibinfo{author}{Fan, J.}, \& \bibinfo{author}{Li, R.} (\bibinfo{year}{2001}).
\newblock \bibinfo{title}{Variable selection via nonconcave penalized
  likelihood and its oracle properties.}
\newblock {\it \bibinfo{journal}{J. Amer. Statist. Assoc.}\/},  {\it
  \bibinfo{volume}{96}\/}, \bibinfo{pages}{1348--1360}.
%Type = Article
\bibitem[{Fan \& Hyndman(2012)}]{FanHyndman2012}
\bibinfo{author}{Fan, S.}, \& \bibinfo{author}{Hyndman, R.~J.}
  (\bibinfo{year}{2012}).
\newblock \bibinfo{title}{Short-term load forecasting based on a
  semi-parametric additive model}.
\newblock {\it \bibinfo{journal}{IEEE Transactions on Power Systems}\/},  {\it
  \bibinfo{volume}{27}\/}, \bibinfo{pages}{134--141}.
%Type = Techreport
\bibitem[{Farrokhabadi et~al.(2021)Farrokhabadi, Browell, Wang, Makonin \&
  Zareipour}]{Farrokhabadi2021}
\bibinfo{author}{Farrokhabadi, M.}, \bibinfo{author}{Browell, J.},
  \bibinfo{author}{Wang, Y.}, \bibinfo{author}{Makonin, W.}, \&
  \bibinfo{author}{Zareipour, H.} (\bibinfo{year}{2021}).
\newblock {\it \bibinfo{title}{Day-Ahead Electricity Demand Forecasting
  Competition: Post-COVID Paradigm}\/}.
\newblock \bibinfo{type}{Technical Report}.
%Type = Article
\bibitem[{Gaillard \& Goude(2016)}]{gaillard2016opera}
\bibinfo{author}{Gaillard, P.}, \& \bibinfo{author}{Goude, Y.}
  (\bibinfo{year}{2016}).
\newblock \bibinfo{title}{opera: Online prediction by expert aggregation}.
\newblock {\it \bibinfo{journal}{URL: https://CRAN. R-project. org/package=
  opera. r package version}\/},  {\it \bibinfo{volume}{1}\/}.
%Type = Inproceedings
\bibitem[{Gaillard et~al.(2014)Gaillard, Stoltz \&
  Van~Erven}]{gaillard2014second}
\bibinfo{author}{Gaillard, P.}, \bibinfo{author}{Stoltz, G.}, \&
  \bibinfo{author}{Van~Erven, T.} (\bibinfo{year}{2014}).
\newblock \bibinfo{title}{A second-order bound with excess losses}.
\newblock In {\it \bibinfo{booktitle}{Conference on Learning Theory}\/} (pp.
  \bibinfo{pages}{176--196}).
%Type = Article
\bibitem[{Genuer et~al.(2015)Genuer, Poggi \& Tuleau-Malot}]{Vsurf}
\bibinfo{author}{Genuer, R.}, \bibinfo{author}{Poggi, J.-M.}, \&
  \bibinfo{author}{Tuleau-Malot, C.} (\bibinfo{year}{2015}).
\newblock \bibinfo{title}{{VSURF: An R Package for Variable Selection Using
  Random Forests}}.
\newblock {\it \bibinfo{journal}{{The R Journal}}\/},  {\it
  \bibinfo{volume}{7}\/}, \bibinfo{pages}{19--33}.
%Type = Article
\bibitem[{Goehry et~al.(2019)Goehry, Goude, Massart \&
  Poggi}]{goehry2019aggregation}
\bibinfo{author}{Goehry, B.}, \bibinfo{author}{Goude, Y.},
  \bibinfo{author}{Massart, P.}, \& \bibinfo{author}{Poggi, J.-M.}
  (\bibinfo{year}{2019}).
\newblock \bibinfo{title}{Aggregation of multi-scale experts for bottom-up load
  forecasting}.
\newblock {\it \bibinfo{journal}{IEEE Transactions on Smart Grid}\/},  {\it
  \bibinfo{volume}{11}\/}, \bibinfo{pages}{1895--1904}.
%Type = Article
\bibitem[{Goude et~al.(2013)Goude, Nedellec \& Kong}]{Goude2013}
\bibinfo{author}{Goude, Y.}, \bibinfo{author}{Nedellec, R.}, \&
  \bibinfo{author}{Kong, N.} (\bibinfo{year}{2013}).
\newblock \bibinfo{title}{{Local Short and Middle term Electricity Load
  Forecasting with semi-parametric additive models}}.
\newblock {\it \bibinfo{journal}{IEEE transactions on smart grid}\/},  {\it
  \bibinfo{volume}{5}\/}, \bibinfo{pages}{440 -- 446}.
%Type = Article
\bibitem[{Khairalla et~al.(2018)Khairalla, Ning, Al-Jallad \&
  El-Faroug}]{khairalla2018short}
\bibinfo{author}{Khairalla, M.~A.}, \bibinfo{author}{Ning, X.},
  \bibinfo{author}{Al-Jallad, N.~T.}, \& \bibinfo{author}{El-Faroug, M.~O.}
  (\bibinfo{year}{2018}).
\newblock \bibinfo{title}{Short-term forecasting for energy consumption through
  stacking heterogeneous ensemble learning model}.
\newblock {\it \bibinfo{journal}{Energies}\/},  {\it \bibinfo{volume}{11}\/},
  \bibinfo{pages}{1605}.
%Type = Article
\bibitem[{Kursa \& Rudnicki(2010)}]{Boruta}
\bibinfo{author}{Kursa, M.~B.}, \& \bibinfo{author}{Rudnicki, W.~R.}
  (\bibinfo{year}{2010}).
\newblock \bibinfo{title}{Feature selection with the {Boruta} package}.
\newblock {\it \bibinfo{journal}{Journal of Statistical Software}\/},  {\it
  \bibinfo{volume}{36}\/}, \bibinfo{pages}{1--13}.
%Type = Article
\bibitem[{Laptev et~al.(2018)Laptev, Yu \& Rajagopal}]{laptev2018applied}
\bibinfo{author}{Laptev, N.}, \bibinfo{author}{Yu, J.}, \&
  \bibinfo{author}{Rajagopal, R.} (\bibinfo{year}{2018}).
\newblock \bibinfo{title}{Applied timeseries transfer learning}, .
%Type = Article
\bibitem[{Meinshausen \& Ridgeway(2006)}]{meinshausen2006quantile}
\bibinfo{author}{Meinshausen, N.}, \& \bibinfo{author}{Ridgeway, G.}
  (\bibinfo{year}{2006}).
\newblock \bibinfo{title}{Quantile regression forests.}
\newblock {\it \bibinfo{journal}{Journal of machine learning research}\/},
  {\it \bibinfo{volume}{7}\/}.
%Type = Article
\bibitem[{Moon et~al.(2020)Moon, Jung, Rew, Rho \& Hwang}]{moon2020combination}
\bibinfo{author}{Moon, J.}, \bibinfo{author}{Jung, S.}, \bibinfo{author}{Rew,
  J.}, \bibinfo{author}{Rho, S.}, \& \bibinfo{author}{Hwang, E.}
  (\bibinfo{year}{2020}).
\newblock \bibinfo{title}{Combination of short-term load forecasting models
  based on a stacking ensemble approach}.
\newblock {\it \bibinfo{journal}{Energy and Buildings}\/},  {\it
  \bibinfo{volume}{216}\/}, \bibinfo{pages}{109921}.
%Type = Techreport
\bibitem[{Obst et~al.(2021{\natexlab{a}})Obst, Ghattas, Cugliari, Oppenheim,
  Claudel \& Goude}]{obst2021transfer}
\bibinfo{author}{Obst, D.}, \bibinfo{author}{Ghattas, B.},
  \bibinfo{author}{Cugliari, J.}, \bibinfo{author}{Oppenheim, G.},
  \bibinfo{author}{Claudel, S.}, \& \bibinfo{author}{Goude, Y.}
  (\bibinfo{year}{2021}{\natexlab{a}}).
\newblock {\it \bibinfo{title}{Transfer Learning for Linear Regression: a
  Statistical Test of Gain}\/}.
\newblock \bibinfo{type}{Technical Report} arXiv:2102.09504.
%Type = Article
\bibitem[{Obst et~al.(2021{\natexlab{b}})Obst, de~Vilmarest \&
  Goude}]{obst2021adaptive}
\bibinfo{author}{Obst, D.}, \bibinfo{author}{de~Vilmarest, J.}, \&
  \bibinfo{author}{Goude, Y.} (\bibinfo{year}{2021}{\natexlab{b}}).
\newblock \bibinfo{title}{Adaptive methods for short-term electricity load
  forecasting during covid-19 lockdown in france}.
\newblock {\it \bibinfo{journal}{IEEE Transactions on Power Systems}\/},  {\it
  \bibinfo{volume}{36}\/}, \bibinfo{pages}{4754--4763}.
%Type = Book
\bibitem[{Olivas et~al.(2009)Olivas, Guerrero, Sober, Benedito \&
  Lopez}]{olivas2009handbook}
\bibinfo{author}{Olivas, E.~S.}, \bibinfo{author}{Guerrero, J. D.~M.},
  \bibinfo{author}{Sober, M.~M.}, \bibinfo{author}{Benedito, J. R.~M.}, \&
  \bibinfo{author}{Lopez, A. J.~S.} (\bibinfo{year}{2009}).
\newblock {\it \bibinfo{title}{Handbook Of Research On Machine Learning
  Applications and Trends: Algorithms, Methods and Techniques - 2 Volumes}\/}.
\newblock \bibinfo{address}{Hershey, PA}: \bibinfo{publisher}{Information
  Science Reference - Imprint of: IGI Publishing}.
%Type = Article
\bibitem[{Pan \& Yang(2010)}]{pan2010survey}
\bibinfo{author}{Pan, S.~J.}, \& \bibinfo{author}{Yang, Q.}
  (\bibinfo{year}{2010}).
\newblock \bibinfo{title}{A survey on transfer learning}.
\newblock {\it \bibinfo{journal}{IEEE Transactions on Knowledge and Data
  Engineering}\/},  {\it \bibinfo{volume}{22}\/}, \bibinfo{pages}{1345--1359}.
%Type = Book
\bibitem[{Schellong(2011)}]{schellong2011energy}
\bibinfo{author}{Schellong, W.} (\bibinfo{year}{2011}).
\newblock {\it \bibinfo{title}{Energy demand analysis and forecast}\/}.
\newblock \bibinfo{publisher}{BoD--Books on Demand}.
%Type = Article
\bibitem[{Smyl(2020)}]{SMYL202075}
\bibinfo{author}{Smyl, S.} (\bibinfo{year}{2020}).
\newblock \bibinfo{title}{A hybrid method of exponential smoothing and
  recurrent neural networks for time series forecasting}.
\newblock {\it \bibinfo{journal}{International Journal of Forecasting}\/},
  {\it \bibinfo{volume}{36}\/}, \bibinfo{pages}{75--85}. \URLprefix
  \url{https://www.sciencedirect.com/science/article/pii/S0169207019301153}.
  \DOIprefix\doi{https://doi.org/10.1016/j.ijforecast.2019.03.017}.
\newblock \bibinfo{note}{M4 Competition}.
%Type = Techreport
\bibitem[{de~Vilmarest \& Goude(2021)}]{de2021state}
\bibinfo{author}{de~Vilmarest, J.}, \& \bibinfo{author}{Goude, Y.}
  (\bibinfo{year}{2021}).
\newblock {\it \bibinfo{title}{State-Space Models Win the IEEE DataPort
  Competition on Post-covid Day-ahead Electricity Load Forecasting}\/}.
\newblock \bibinfo{type}{Technical Report} arXiv:2110.00334.
%Type = Article
\bibitem[{Wang et~al.(2019)Wang, Chen, Hong \& Kang}]{Wang2019}
\bibinfo{author}{Wang, Y.}, \bibinfo{author}{Chen, Q.}, \bibinfo{author}{Hong,
  T.}, \& \bibinfo{author}{Kang, C.} (\bibinfo{year}{2019}).
\newblock \bibinfo{title}{Review of smart meter data analytics: Applications,
  methodologies, and challenges}.
\newblock {\it \bibinfo{journal}{IEEE Transactions on Smart Grid}\/},  {\it
  \bibinfo{volume}{10}\/}, \bibinfo{pages}{3125--3148}.
%Type = Book
\bibitem[{Wood(2017)}]{GAMbook}
\bibinfo{author}{Wood, S.} (\bibinfo{year}{2017}).
\newblock {\it \bibinfo{title}{Generalized Additive Models: An Introduction
  with R}\/}.
\newblock (\bibinfo{edition}{2nd} ed.).
\newblock \bibinfo{publisher}{Chapman and Hall/CRC}.
%Type = Book
\bibitem[{Wood(2006)}]{wood2006generalized}
\bibinfo{author}{Wood, S.~N.} (\bibinfo{year}{2006}).
\newblock {\it \bibinfo{title}{Generalized additive models: an introduction
  with R}\/}.
\newblock \bibinfo{publisher}{chapman and hall/CRC}.
%Type = Article
\bibitem[{Xenochristou \& Kapelan(2020)}]{xenochristou2020ensemble}
\bibinfo{author}{Xenochristou, M.}, \& \bibinfo{author}{Kapelan, Z.}
  (\bibinfo{year}{2020}).
\newblock \bibinfo{title}{An ensemble stacked model with bias correction for
  improved water demand forecasting}.
\newblock {\it \bibinfo{journal}{Urban Water Journal}\/},  {\it
  \bibinfo{volume}{17}\/}, \bibinfo{pages}{212--223}.
%Type = Article
\bibitem[{Zhai \& Chen(2018)}]{zhai2018development}
\bibinfo{author}{Zhai, B.}, \& \bibinfo{author}{Chen, J.}
  (\bibinfo{year}{2018}).
\newblock \bibinfo{title}{Development of a stacked ensemble model for
  forecasting and analyzing daily average pm2. 5 concentrations in beijing,
  china}.
\newblock {\it \bibinfo{journal}{Science of The Total Environment}\/},  {\it
  \bibinfo{volume}{635}\/}, \bibinfo{pages}{644--658}.
%Type = Article
\bibitem[{Zhuang et~al.(2020)Zhuang, Qi, Duan, Xi, Zhu, Zhu, Xiong \&
  He}]{zhuang2020comprehensive}
\bibinfo{author}{Zhuang, F.}, \bibinfo{author}{Qi, Z.}, \bibinfo{author}{Duan,
  K.}, \bibinfo{author}{Xi, D.}, \bibinfo{author}{Zhu, Y.},
  \bibinfo{author}{Zhu, H.}, \bibinfo{author}{Xiong, H.}, \&
  \bibinfo{author}{He, Q.} (\bibinfo{year}{2020}).
\newblock \bibinfo{title}{A comprehensive survey on transfer learning}.
\newblock {\it \bibinfo{journal}{Proceedings of the IEEE}\/},  {\it
  \bibinfo{volume}{109}\/}, \bibinfo{pages}{43--76}.
%Type = Techreport
\bibitem[{Ziel(2021)}]{ziel2021smoothed}
\bibinfo{author}{Ziel, F.} (\bibinfo{year}{2021}).
\newblock {\it \bibinfo{title}{Smoothed Bernstein Online Aggregation for
  Day-Ahead Electricity Demand Forecasting}\/}.
\newblock \bibinfo{type}{Technical Report} arXiv:2107.06268.

\end{thebibliography}

\appendix
\section{Variable selection for electricity load forecasting during the first Covid-19 lockdown}
\label{app:selectionVariables}
We allow the RF to use many variables for their predictions, including usual calendar and weather variables, as well as mobility data, containment index, and estimated GAM effects, without knowing a priori which ones will be relevant to predict electricity consumption during the pandemic period. It is reasonable to think that including all covariates might be detrimental to the prediction, given the high correlations between some variables and the small number of observations available to train the model, especially in the early days of the lockdown. 

\paragraph{Variable selection for the RF}
We allow the RF to use many variables for their predictions, including usual calendar and weather variables, as well as mobility data, containment index, and estimated GAM effects, without knowing a priori which ones will be relevant to predict electricity consumption during the pandemic period. It is reasonable to assume that including all covariates might be detrimental to the prediction, given the high correlations between some variables and the small number of observations available to train the model, especially in the early days of lockdown. We want to take advantage of the fact that the number of observations increases rapidly, by repeating the variable selection operation several times during the pandemic period, in order to be able to enrich the model if necessary. Moreover, we expect that the relevant variables might differ from one region to the other, and so we want to perform the variable selection region by region.

To do so, we select every week the variables to be used to train the RF for the following week's forecasts for a  given region. Feature selection in RF is an ongoing field of research. State-of-the-art methods rely on the ranking of variable importance measure, such as VSURF \citep{Vsurf}, or on permutation of variables, such as Boruto \citep{Boruta}.  These methods suffer from their important computational cost: VSURF, in particular, is too slow to be used in our context of numerous variable selection operations. We suggest an alternative approach to determine the relevant covariates, using the technics developed for variable selection in linear regression. More precisely, we fit a linear model to predict the residuals of the GAM during the pandemic period using a LASSO penalty. Without prior knowledge of the number of covariates necessary to accurately forecast electricity load, we design three models corresponding  to different numbers of covariates. More precisely, we fit a LASSO with an ad-hoc penalty to select respectively 5 and 10 covariates. This variable selection step is repeated every week. Finally, the predictions of the RF taking these 5 (respectively 10) covariates as input are combined with that of the RF taking all covariates as input using an expert aggregation method.

Before implementing this method for all regions and all quantiles, we evaluate its interest in predicting the 0.5 quantile of the national load using the available national data. We compare the MAPE of the predictor obtained using Boruto, Lasso variable selection with an aggregation step, and of the full model with 16 variables. The MAPE for the lockdown and post-lockdown periods are presented in Table \ref{table:VarSel}.

\begin{table}[!h] \center
\begin{tabular}{c|c|c}
    Selection method & \small{2020/03/16-2020/05/11} & \small{2020/05/11-2020/09/17}\\
    \hline
    Lasso with aggregation & 2.41 \%  & 1.06 \%\\
    \hline
    Boruta &  2.41 \%& 1.03 \%\\
    \hline
    Full model & 2.41 \% & 1.03 \% \\
\end{tabular}
\caption{\label{table:VarSel}Mean Absolute Percentage Error of the stacked GAM and RF models.}
\end{table}

The preliminary results indicate that variable selection only marginally affects the performance of the RF. This underlines the robustness of the RF against a high dimension of the inputs, even when trained on relatively small data sets. We therefore consider the RF obtained using the full models, taking as input all covariates and estimated GAM effects.

\section{Analysis of our method for electricity load forecasting during the first Covid-19 lockdown} \label{app:covid}
\subsection{Analysis of the stacked GAM and RF}

\begin{figure}[h!]
  \begin{center}
    \subfloat[Average importance of types of variables in \newline the staked RF at the national level.]{
      \includegraphics[width=0.42\textwidth]{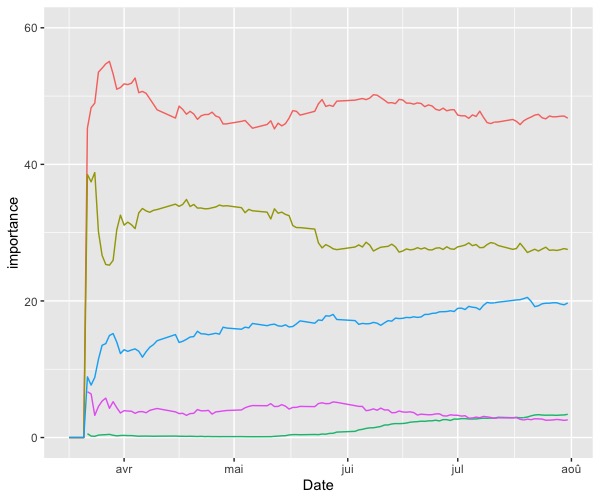}
      \label{subfig:groupe_France}
                         }
    \subfloat[Average importance of types of variables in the staked RF common to all regions and the national level.]{
     \includegraphics[width=0.53\textwidth]{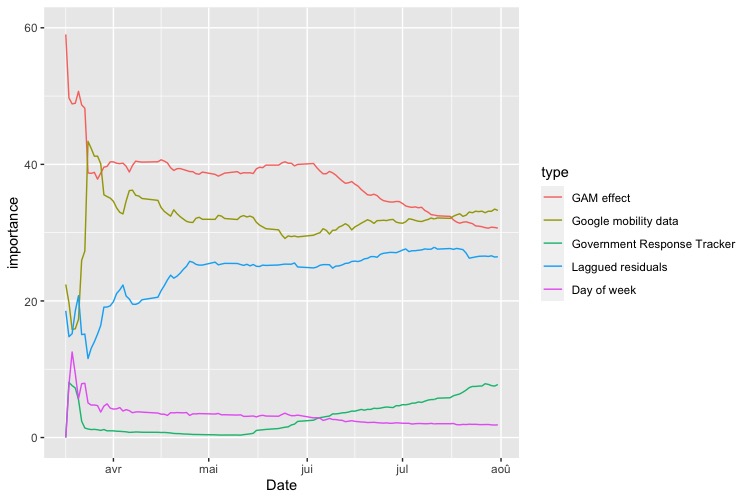}
     \label{subfig:groupe_common}
                         }

      \subfloat[Average importance of mobility measures in \newline the staked RF at the national level.]{
      \includegraphics[width=0.42\textwidth]{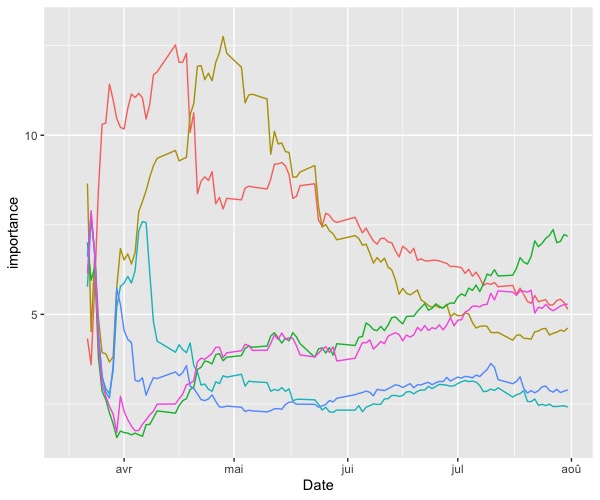}
      \label{subfig:google_France}
                         }
    \subfloat[Average importance of types of mobility measures in the staked RF \ \ common to all regions and the national level.]{
     \includegraphics[width=0.53\textwidth]{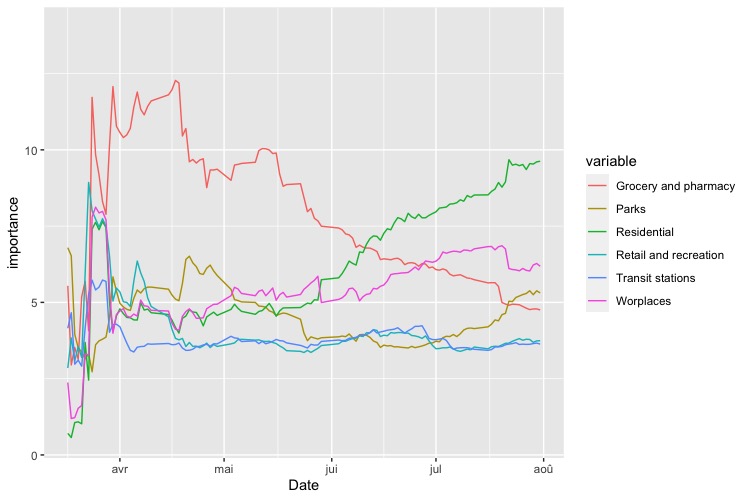}
     \label{subfig:google_common}
                         }
    
    \caption{Evolution of the importance of the types of variables (top) and the mobility measures (bottom) in the RF trained on GAM residuals at the national level (left), and on GAM residuals for all regions and at  the national level. }
    \label{fig:importance}
  \end{center}
\end{figure}

Figures \ref{subfig:groupe_France} and \ref{subfig:groupe_common} present the evolution of the average importance (for the different half-hours) of the variables for the stacked RF trained respectively on residuals at the national level, and on all residuals. The importances of the different variables for a given model are normalized so that their sum remains constant during the pandemic period, and equal to 100. More precisely, denoting by $I_{v,t}$ and $I_{v,t}^{normalized}$ respectively the importance and normalized  importance of variable $v$ at time $t$, we have at any time $t$
\begin{equation*}
    I_{v,t}^{normalized} = 100*\frac{I_{v,t}}{\sum_{variables\ v'} I_{v',t}}
\end{equation*}
We group the variables into 5 categories: the GAM effects, the measures of mobility, the government response tracker, the lagged residuals, and the day of the week. The importance of a group is simply the sum of the importances of the variables in this group. The importance of the mobility measures is detailed in Figures \ref{subfig:google_France} and \ref{subfig:google_common}.

\begin{figure}[h!]
  \begin{center}
    \includegraphics[width=\textwidth]{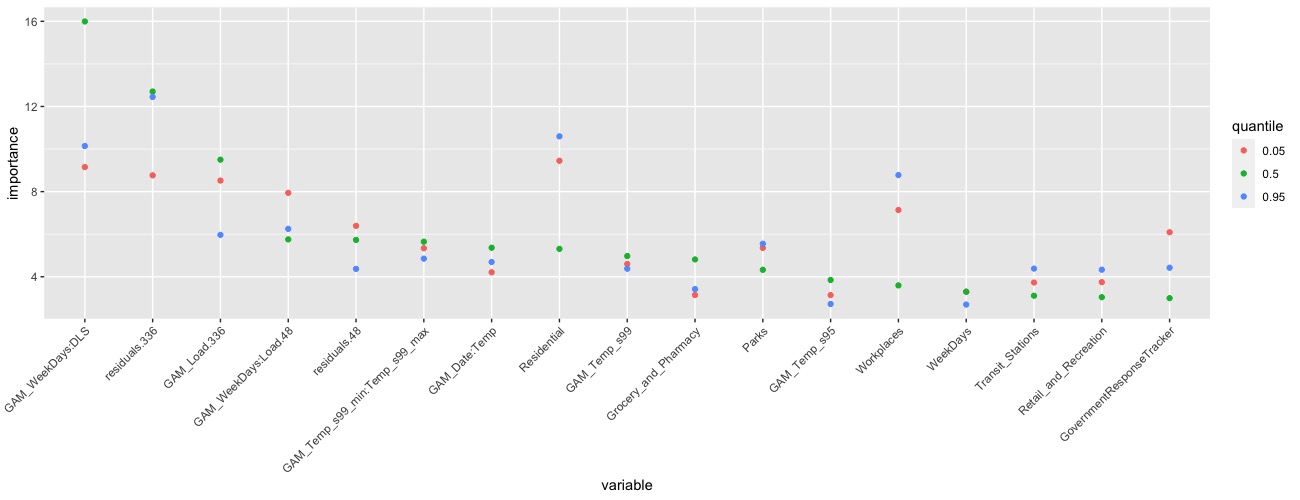}
                         
\caption{\label{fig:quantile_importance}Importance of the variables in the stacked RF predicting the quantiles 0.05, 0.5 and 0.95 of the GAM residuals at the national level. The variables ``${\rm GAM}_X$" denotes the GAM effect corresponding to variable ``X".}
 \end{center}
\end{figure}

\begin{figure}[h!]
  \begin{center}

    \subfloat[Accumulated Local Effects of the relative \newline frequentation of residential places.]{
      \includegraphics[width=0.45\textwidth]{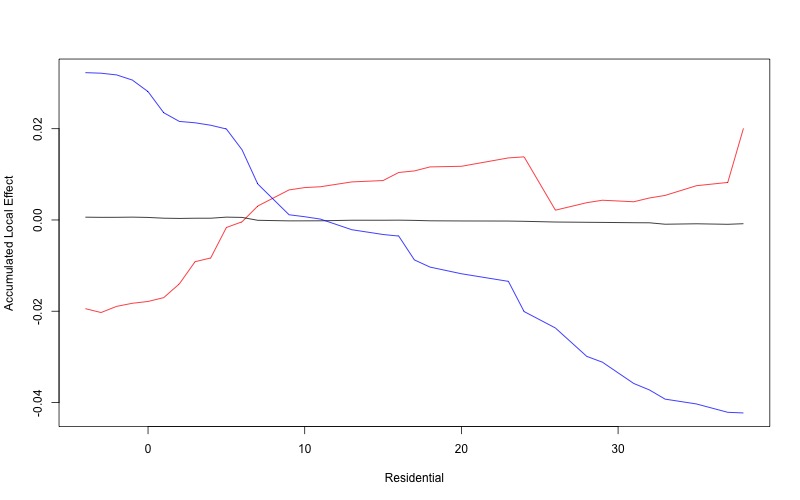}
                         }
    \subfloat[Accumulated Local Effects of the measure of \newline relative frequentation of workplaces.]{
      \includegraphics[width=0.45\textwidth]{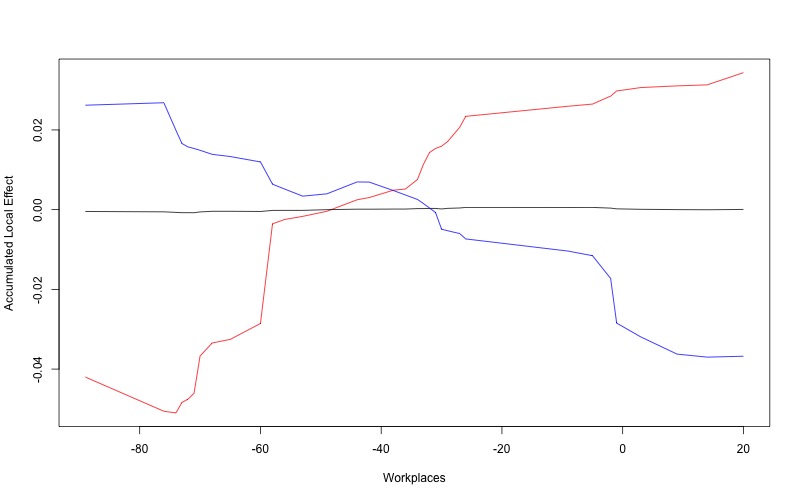}
                         }
\caption{\label{fig:ALE}Accumulated Local Effects of the measure of relative frequentation of residential places (left) and workplaces (right) for the RF at the national level predicting the quantiles 0.05 (red), 0.5 (black), and 0.95 (blue).}
 \end{center}
\end{figure}

We note that the effects of the GAM are among the most important covariates for predicting the GAM residuals. Using these effects as covariates allows to transfer information on the impact of weather and calendar variables learned on the large dataset of pre-pandemic observations. We observe a change in the importance of the different types of variables after the end of the lockdown, indicating that the RF is able to account for a relative change in electricity consumption patterns. As time passes and the size of the training set for the RF increases, relevant variables such as the Government Response Tracker, or relative occupation of residence, and grocery and pharmacies, become more important for the prediction. Conversely, spurious variables (for example, the relative frequentation of parks, highly correlated with weather) are discarded as unimportant. Interestingly, the common RF trained on residuals across all regions detects these relevant variables more quickly than the individual RF trained solely on residuals at the national level. This highlights the interest of multi-task learning in this sparse data context.

We also investigate the relative importance of the variables in the stacked RF predicting the different quantiles. More precisely, we consider the stacked RF trained on the residuals at the national level for the pandemic period. We compute an importance measure of a given variable as the average increase in error in term of the pinball loss corresponding to a given quantile when the values of this variable are permuted at random (the error is computed over the training set). The importances of the different variables are normalized so that their sum is equal to 100. We compare the importance of the variables for predicting the 0.05, 0.5, and 0.95 quantiles in Figure \ref{fig:quantile_importance}. Variables important for predicting one quantile tend to be important for predicting the other quantiles. However, this is not the case for all variables: for example, the normalized load for the relative frequentation of residential place and workplaces have an outstanding importance for predicting the 0.05 and 0.95 quantiles. These variables have a very high (negative) correlation: the frequentation of workplaces is very low during the lockdown period, and remains relatively low in the post-lockdown period during weekdays; on the opposite, the frequentation of residential places is high during the lockdown period and remains relatively high in the post-lockdown period during weekdays. Looking at the Accumulated Local Effects of these variables, plotted in Figure \ref{fig:ALE}, we see that they have a much larger impact on the prediction of the two extreme quantiles than on that of the median. However, we expect their effects to partially cancel each other out because of the correlation between these variables.

\subsection{Analysis of online aggregation}\label{subsec:aggregationCovid}
\begin{figure}[!h]
  \begin{center}
    \includegraphics[width=\textwidth]{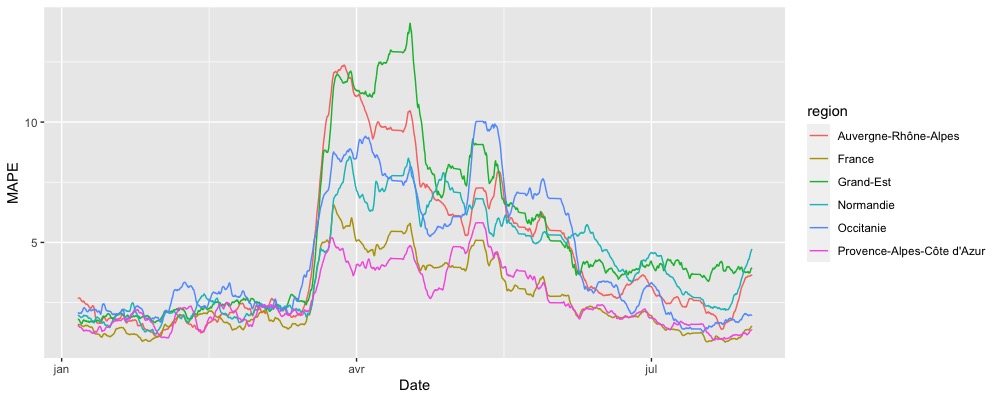}
 \end{center}
 \caption{
    \label{fig:MAPE_GAM_region}Weekly averaged MAPE of the normalized GAM for the regions Auvergne-Rhône-Alpes, Grand-Est, Normandie, Occitanie, and at the national level.}
\end{figure}

Our results indicate that online aggregation is an efficient way to take into account information available at a finer scale. Note that the regional GAM have in average errors much larger than that of the GAM at the national level, as illustrated in Figure \ref{fig:MAPE_GAM_region}, due to larger fluctuations present at the finer scale. Interestingly, aggregating these low-accuracy models allows to obtain better performances than that of the GAM at the national level, even in the pre-pandemic period, as indicated by our results in Table \ref{table:MAPE}.

\begin{figure}[!h]
  \begin{center}
    \subfloat[\textbf{Weights of the regional and national experts in the prediction at the national level.} Red: sum of the weights of the quantile and GAM experts by region, in the aggregation targeting the national load using the full disaggregated approach. Green: weights of the regional experts, and for the national level in the aggregation targeting the national load using the scaled hierarchical approach. Blue: true proportion of the national electricity load consumed by the region. ]{
    \includegraphics[width=0.9\textwidth]{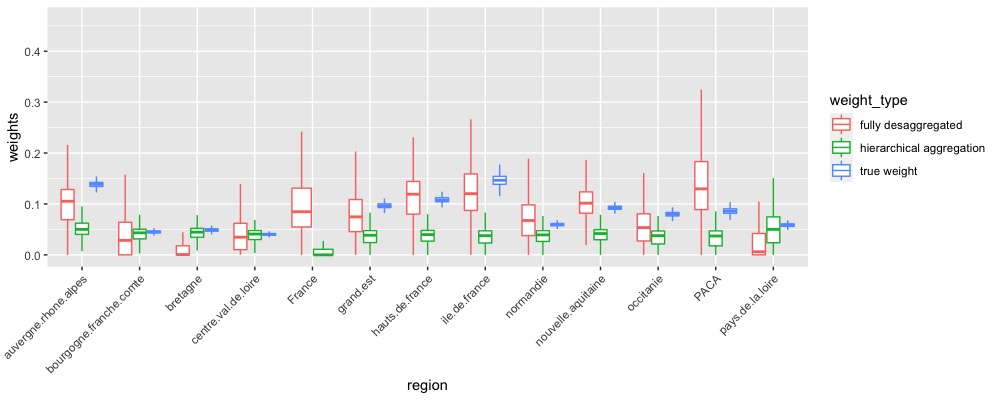}
    \label{fig:weights_regions}
                         }
                         
    \subfloat[\textbf{Weights of the quantile and GAM experts in the prediction at the national level.} Weights of the quantile experts and the GAM expert in the aggregation targeting the national load using a full disaggregated approach (red), a hierachical aggregation approach (green), and a vectorial aggregation approach (blue).]{
     \includegraphics[width=0.9\textwidth]{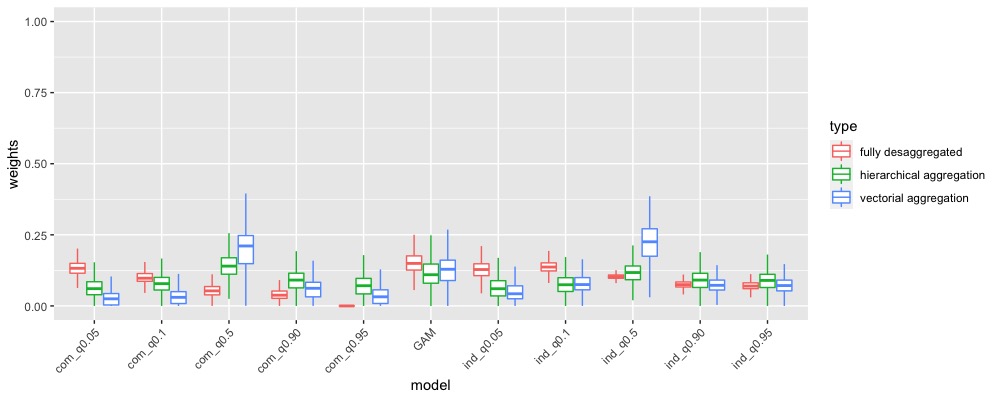}
     \label{subfig:weight_quantile}
                         }
 \end{center}

\end{figure}

\begin{figure}[!h]
  \begin{center}
    \subfloat[Weights of the GAM and quantile experts in \newline the  first step of hierarchical aggregation, targeting \newline the national load at 7:30 pm.]{
    \includegraphics[width=0.45\textwidth]{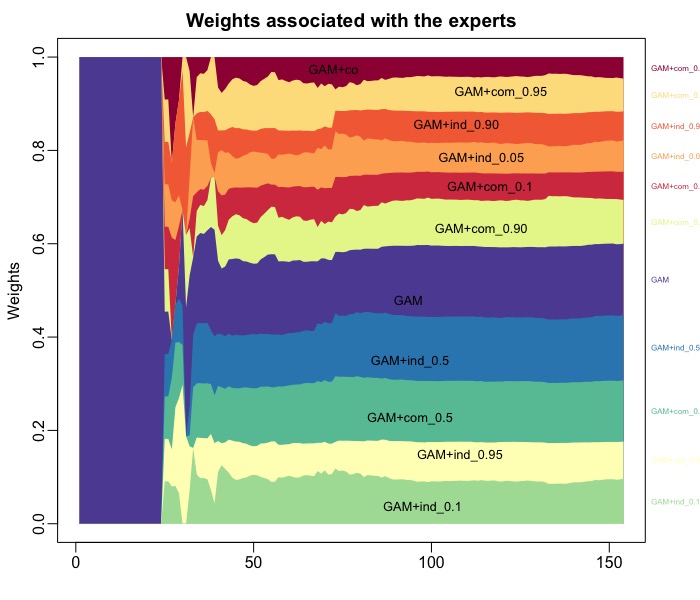}
    \label{subfig:dynamical_weights_France}
                         }
    \subfloat[Evolution of the weights of the GAM and quantile \newline experts in the vectorial aggregation at 7:30 pm.]{
     \includegraphics[width=0.45\textwidth]{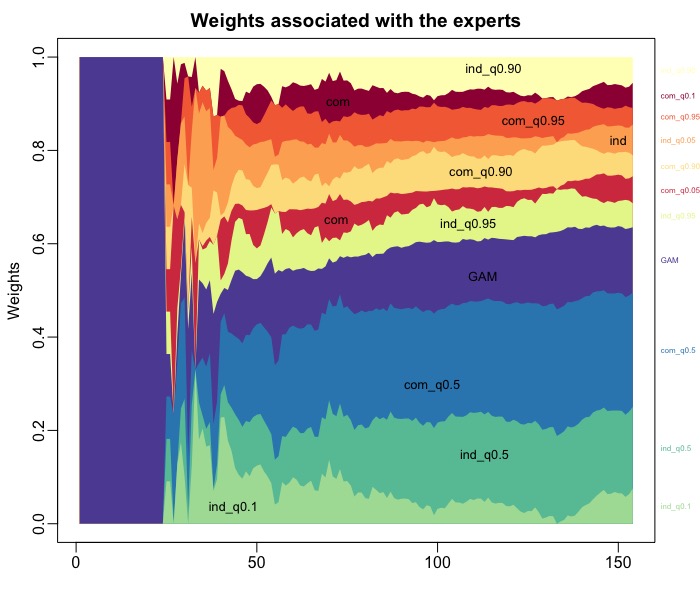}
     \label{subfig:dynamical_weights_vectorial}
                         }
 \end{center}
 \caption{\label{fig:dynamical_weights}Evolution of the weights of the quantile stacked GAM-RF experts and the GAM expert in the prediction of national load.}
\end{figure}

The fact that scaled and unscaled hierarchical aggregation obtain similar performances is somewhat counterintuitive, given that in the scaled model the aggregation must learn the contribution of the different regions to the national consumption. Looking at the distribution of the weights of the regions in the scaled hierarchical aggregation presented in Figure \ref{fig:weights_regions}, we note that the weights do not correspond to the proportion of electricity consumed by the regions (for example, regions with low true weights such as Provence-Alpes-Côte d'Azur may receive more weight in the aggregation than regions with high true weights, such as Île-de-France). Moreover, the weights in the aggregation typically exhibit much more flexibility than the true weights: this phenomenon is all the more striking in the fullydisaggregated model. The high variability of the weights suggests that some of the models considered are fairly interchangeable. The fact that the scaled hierarchical aggregation outperforms its unscaled counterpart both in the pre-pandemic and in the post-lockdown period suggests that the flexibility provided by the second layer of aggregation used in the scaled model compensates the lack of knowledge of the relative contribution of the different regions. 

We see in Figure \ref{subfig:weight_quantile} that all quantiles and GAM experts contribute to the prediction, both in the fully disaggregated model and in the hierarchical aggregation. By contrast, the vectorial aggregation gives a predominant weight to the GAM and median staked RF experts, which appears as the most relevant experts across all regions. Figure \ref{fig:dynamical_weights} shows the weights given by an aggregation predicting the national load using only the national experts, and the weights given by the aggregation. Day 25 corresponds to the first day of the pandemic period; before that day only the GAM forecast is available to the aggregation. We note that the weights in the vectorial aggregation are highly unstable during the beginning of the lockdown. The performance of the vectorial aggregation during this period is worst than that of all other aggregation models, and than that of the stacked RF predicting the median of residuals. This behavior mirrors the fact that the impact of the pandemic strongly differs from one region to another, as is shown in Figure \ref{fig:MAPE_GAM_region}. On the other hand, vectorial aggregation achieves the best performance during the post-lockdown period, and appears as a promising approach to predicting consumption under normal circumstances.

\end{document}